\documentclass[aps,prd,twocolumn,tightenlines,preprintnumbers,10pt,nofootinbib,floatfix]{revtex4-2}
\pdfoutput=1

\usepackage{graphicx}
\usepackage{amssymb}
\usepackage{amsmath}
\usepackage[shortlabels]{enumitem}
\usepackage[caption=false]{subfig}
\usepackage{slashed}
\usepackage{appendix}
\usepackage{booktabs}
\usepackage{listings}
\usepackage{hyperref}

\usepackage[section]{placeins}
\usepackage{bm}
\usepackage{bbm}
\usepackage{bbold}
\usepackage{multirow}

\usepackage{amsthm}

\newcommand{\diasize}{2cm}
\newcommand{\renormdiasize}{2cm}

\newcommand{\Tr}{\text{Tr}}
\newcommand{\TrC}{\text{Tr}_{\text{C}}}
\newcommand{\TrD}{\text{Tr}_{\text{D}}}

\newcommand{\MS}{\overline{\mathrm{MS}}}

\usepackage{simpler-wick}
\usepackage[compat=1.0.0]{tikz-feynman}   

\DeclareRobustCommand\circled[1]{\tikz[baseline=(char.base)]{\node[shape=circle,draw,inner sep=2pt] (char) {#1};}}

\newcommand{\eff}{{\mathrm{ef}\hspace{-0.2mm}\mathrm{f}}}
\usepackage{tcolorbox}

\begin{document}

\title{Neutrinoless Double Beta Decay from Lattice QCD: The Short-Distance $\pi^-\rightarrow\pi^+ e^- e^-$ Amplitude}
\author{W. Detmold, W. I. Jay, D. J. Murphy, P. R. Oare, P. E. Shanahan}
\affiliation{Center for Theoretical Physics, Massachusetts Institute of Technology, Boston, MA 02139, USA and \\ The NSF AI Institute for Artificial Intelligence and Fundamental Interactions}

\date{\today}
\preprint{MIT-CTP/5414}

\begin{abstract}
This work presents a determination of potential short-distance contributions to the unphysical $\pi^-\rightarrow\pi^+ e^- e^-$ decay through lattice QCD calculations. The hadronic contributions to the transition amplitude are described by the pion matrix elements of five Standard Model Effective Field Theory operators, which are computed on five ensembles of domain-wall fermions with $N_f = 2 + 1$ quark flavors with a range of heavier-than-physical values of the light quark masses. The matrix elements are extrapolated to the continuum, physical light-quark mass, and infinite volume limit using a functional form derived in chiral Effective Field Theory ($\chi\mathrm{EFT}$). This extrapolation also yields the relevant low-energy constants of $\chi\mathrm{EFT}$, which are necessary input for $\chi\mathrm{EFT}$ calculations of neutrinoless double beta decay of nuclei.
\end{abstract}

\maketitle

\section{Introduction}
 
Neutrinoless double beta ($0\nu\beta\beta$) decay, if observed, would unambiguously reveal the existence of physics beyond the Standard Model (BSM)~\cite{Dolinski:2019nrj}. In particular, it would imply that the difference between baryon number and lepton number $(B - L)$ is not a fundamental symmetry of the universe~\cite{PhysRevD.25.774}, and would prove that the neutrino is a Majorana particle~\cite{Majorana2006}. Moreover, observation of $0\nu\beta\beta$ decay would provide additional information about the matter-antimatter asymmetry in the universe~\cite{Deppisch:2017ecm}, which may help to explain baryogenesis and further constrain the neutrino masses~\cite{Vergados:2016hso}.

As such, experiments are underway worldwide to search for $0\nu\beta\beta$ decay, the most sensitive of which study $^{76}\mathrm{Ge}$ and $^{136}\mathrm{Xe}$ and constrain the half-lives of $0\nu\beta\beta$ decay in each isotope to be greater than  $10^{26}\;\mathrm{years}$~\cite{Menendez:2017fdf,https://doi.org/10.48550/arxiv.2202.01787,PhysRevLett.111.122503, Gando:2020cxo,RevModPhys.80.481}. Understanding the implication of these constraints for possible BSM physics scenarios requires input in the form of nuclear matrix elements (NMEs); which NMEs are relevant depends on the underlying mechanism of $0\nu\beta\beta$ decay. These mechanisms can be broadly divided into two categories: long-distance mechanisms, in which the decay is induced by a non-local interaction mediated by a light particle of mass much less than the hadronic scale~\cite{GRIFOLS1996563,Kotila:2021xgw}; and short-distance mechanisms, in which the decay is mediated by a heavy particle that can be integrated out in Effective Field Theory (EFT) to generate contact interactions~\cite{Graf:2018ozy,PhysRevD.98.095023}. In extensions of the Standard Model, long-distance mechanisms are typically assumed to be generated by the dimension-5 Weinberg operator, in which the mediating particle is generally a light Majorana neutrino (although other scenarios have been considered)~\cite{PhysRevD.102.095016,Cirigliano:2017djv,Rodejohann:2011mu,Bilenky:2014uka}, while short-distance mechanisms are described by operators of dimension greater than or equal to 9~\cite{Weinberg:1979sa}. The dominant mechanism of $0\nu\beta\beta$ decay will determine the scale $\Lambda_\mathrm{LNV}$ at which lepton-number violating physics is observed. In particular, if $0\nu\beta\beta$ decay is primarily described by a long-distance mechanism, then $\Lambda_\mathrm{LNV}\gg 1\;\mathrm{TeV}$~\cite{MOHAPATRA1999376}, while if $0\nu\beta\beta$ decay is primarily described by a short-distance mechanism, $\Lambda_\mathrm{LNV}\sim 1\;\mathrm{TeV}$~\cite{DellOro:2016tmg,Cirigliano:2018yza}. Both cases must be understood in order to draw conclusions about the underlying BSM physics from any experimental detection of $0\nu\beta\beta$ decay.

Calculations of long and short-distance $0\nu\beta\beta$ decay matrix elements have been performed with nuclear many-body methods~\cite{DellOro:2016tmg,Engel:2016xgb}. These techniques are currently the only theoretical methods which can provide insight into $0\nu\beta\beta$ decay in nuclear isotopes which are experimentally relevant. 
The requisite NMEs for the long and short-distance $0\nu\beta\beta$ decay of $^{48}\mathrm{Ca}$, $^{76}\mathrm{Ge}$ and $^{136}\mathrm{Xe}$ have been computed, although large model dependence in the calculated NMEs remains a challenge for these techniques~\cite{Menendez:2017fdf,Jokiniemi:2021qqv,https://doi.org/10.48550/arxiv.2202.01787,Barea:2015kwa, PhysRevD.90.096010}. To improve these calculations, connection to the Standard Model is required. 

Lattice quantum chromodynamics (LQCD) is the only known method with which to compute NMEs directly from the underlying quark and gluon degrees of freedom. However, current LQCD calculations of nuclei suffer from a signal-to-noise problem~\cite{PARISI1984203, Lepage:1989hd} and a factorial increase in the number of quark contractions with atomic number~\cite{PhysRevD.87.114512}, which make calculations of phenomologically relevant nuclei impractical in the absence of new algorithms and approaches. Instead of direct computation of large nuclei, recent work uses EFT~\cite{Savage:1998yh,PhysRevD.68.034016,Graesser:2016bpz,Cirigliano:2020yhp,Cirigliano:2018yza,Cirigliano:2017djv} to relate LQCD calculations of simpler processes such as the unphysical mesonic transition $\pi^-\rightarrow\pi^+ e^- e^-$ and the two-nucleon $0\nu\beta\beta$ decay  $n^0n^0\rightarrow p^+ p^+ e^- e^-$ to nuclear $0\nu\beta\beta$ decay. Studies of the $\pi^-\rightarrow\pi^+ e^- e^-$ transition in particular do not incur the technical challenges faced by LQCD calculations of nuclei. The long-distance pion matrix elements have been computed directly using LQCD with a domain-wall fermion action~\cite{Detmold:2020jqv,PhysRevD.100.094511}. The associated short-distance pion matrix elements have been calculated from LQCD input with two approaches: relating the desired matrix elements to kaon-mixing matrix elements, assuming $SU(3)$ chiral symmetry~\cite{Cirigliano:2017ymo}; and computing the pion matrix elements directly using LQCD with a mixed action~\cite{PhysRevLett.121.172501}.

This work presents a direct LQCD computation of the $\pi^-\rightarrow\pi^+ e^- e^-$ matrix elements of the leading short-distance (dimension-9) operators, performed for $m_e = 0$ and at threshold. This calculation uses domain-wall fermions, as their chiral symmetry properties yield matrix elements that have a simple renormalization structure. There is a mild tension between the results of the present calculation and the previous mixed-action LQCD calculation of the same matrix elements in Ref.~\cite{PhysRevLett.121.172501}, which may be due to the differences in the action used in each calculation. 
The ensembles used in this calculation are the same as those used in the first lattice computation of the long-distance $\pi^-\rightarrow\pi^+ e^- e^-$ amplitude mediated by light Majorana neutrino exchange~\cite{Detmold:2020jqv}. As such, both the long and short-distance contributions to $\pi^-\rightarrow\pi^+ e^- e^-$ have now been computed in a consistent framework, allowing conclusions to be drawn regarding the relative importance of the two potential contributions, as discussed in Section~\ref{sec:conclusion}.

The remainder of this paper is organized as follows. Section~\ref{sec:sd_matrix_elems} details the EFT framework for the short-distance $\pi^-\rightarrow\pi^+e ^- e^-$ decay and the LQCD calculation of the hadronic part of the transition amplitude. Section~\ref{sec:chiral_extrap} describes the procedure used to extrapolate the renormalized LQCD matrix elements to the physical point using a model based on chiral EFT ($\chi$EFT), and presents results for the extrapolated matrix elements and the extracted $\chi\mathrm{EFT}$ low-energy constants (LECs). Section~\ref{sec:conclusion} summarizes the results and presents an outlook.

\section{Short-distance matrix elements}
\label{sec:sd_matrix_elems}

\subsection{Short-distance operators}

\begin{figure*}[!t]
	\centering
    \subfloat[$\pi\pi$ vertex.]{
	\begin{tikzpicture}
	  \begin{feynman}
	    \vertex (a1) at (0, 2cm);
	    \vertex (a3) at (0, 0);
	    \vertex (a2) at (0, -2cm);
	    \vertex (p1) at (2cm, 2cm) { \( p^+ \) };
	    \vertex (p2) at (2cm, -2cm) { \( p^+ \) };
	    \vertex (n1) at (-2cm, 2cm) { \( n^0 \) };
	    \vertex (n2) at (-2cm, -2cm) { \( n^0 \) };
	    \vertex (e1) at (2cm, 0.5cm) {\( e^- \)};
	    \vertex (e2) at (2cm, -0.5cm) {\( e^- \)};
	    \node[blob] at (0, 0) (a) {};
	    \diagram* {
	    (n2) -- [fermion] (a2) -- [fermion] (p2),
	    (n1) -- [fermion] (a1) -- [fermion] (p1),
	    (a1) -- [scalar] (a) -- [scalar] a2,
	    (a) -- [fermion] (e1),
	    (a) -- [fermion] (e2)
	    };
	  \end{feynman}
	\end{tikzpicture}
	\label{fig:pion_decay}
    }
	~
    \subfloat[$NN$ vertex.]{
	\begin{tikzpicture}
	  \begin{feynman}
	    \vertex (a1) at (0, 0);
	    \node[blob] at (0, 0) (a) {};
	    \vertex (p1) at (\diasize, \diasize) { \( p^+ \) };
	    \vertex (p2) at (\diasize, -\diasize) { \( p^+ \) };
	    \vertex (n1) at (-\diasize, \diasize) { \( n^0 \) };
	    \vertex (n2) at (-\diasize, -\diasize) { \( n^0 \) };
	    \vertex (e1) at (\diasize, 0.5cm) {\( e^- \)};
	    \vertex (e2) at (\diasize, -0.5cm) {\( e^- \)};

	    \diagram* {
	    (n2) -- [fermion] (a) -- [fermion] (p1),
	    (n1) -- [fermion] (a) -- [fermion] (p2),
	    (a) -- [fermion ] (e1),
	    (a) -- [fermion] (e2)
	    };
	  \end{feynman}
	\end{tikzpicture}
	\label{fig:nn_vertex}
    }
	~
    \subfloat[$\pi N$ vertex.]{
	\begin{tikzpicture}
	  \begin{feynman}
	    \vertex (a1) at (0, \diasize);
	    \vertex (a2) at (0, -\diasize);
	    \vertex (p1) at (\diasize, \diasize) { \( p^+ \) };
	    \vertex (p2) at (\diasize, -\diasize) { \( p^+ \) };
	    \vertex (n1) at (-\diasize, \diasize) { \( n^0 \) };
	    \vertex (n2) at (-\diasize, -\diasize) { \( n^0 \) };
	    \vertex (e1) at (\diasize, 0.5cm) {\( e^- \)};
	    \vertex (e2) at (\diasize, -0.5cm) {\( e^- \)};
	    \node[blob] at (0, -\diasize) (a) {};

	    \diagram* {
	    (n2) -- [fermion] (a) -- [fermion] (p2),
	    (n1) -- [fermion] (a1) -- [fermion] (p1),
	    (a1) -- [scalar] (a),
	    (a) -- [fermion] (e1),
	    (a) -- [fermion] (e2)
	    };
	  \end{feynman}
	\end{tikzpicture}
	}
    \caption{Diagrams illustrating short-distance contributions to the $n^0 n^0 \rightarrow p^+ p^+ e^- e^-$ $0\nu\beta\beta$ decay in $\chi$EFT. The solid lines denote nucleons or electrons and the dotted lines denote pions. The hatched circles represent EFT operators built from hadronic fields, which at LO for the $\pi\pi$ vertex diagram, Fig.~(\ref{fig:pion_decay}), are determined by $\mathcal O_k^\chi$ in Eq.~\eqref{eq:chipt_lagrangian}. The $\pi\pi$ (Fig.~(\ref{fig:pion_decay})) and $NN$ (Fig.~(\ref{fig:nn_vertex})) diagrams are the LO $\chi\mathrm{EFT}$ contributions to $n^0 n^0 \rightarrow p^+ p^+ e^- e^-$.}
    \label{fig:0nubb_modes}
\end{figure*}
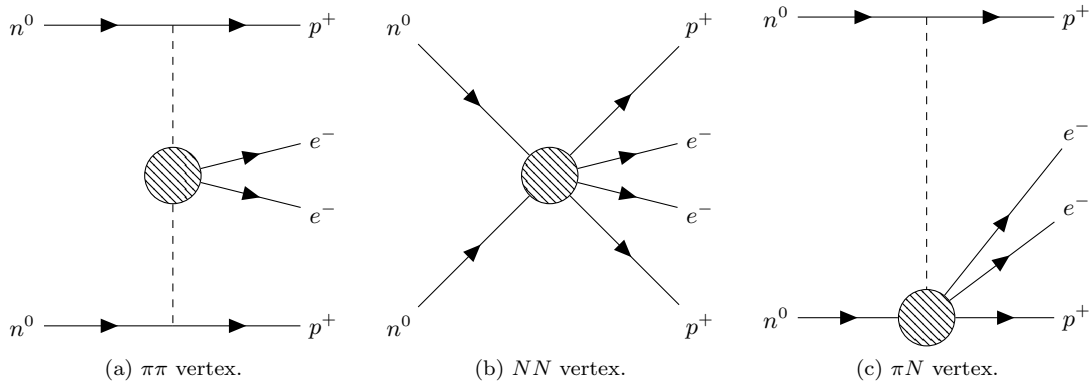

In the Standard Model EFT (SMEFT) framework, the Standard Model enters as the renormalizable sector of a non-renormalizable theory~\cite{Brivio:2017vri}. Potential short-distance contributions to $\pi^-\rightarrow\pi^+ e^- e^-$ are induced by physics at the scale $\Lambda_\mathrm{LNV}\gtrsim v$, where $v = 247\;\mathrm{GeV}$ is the electroweak scale set by the Higgs vacuum expectation value, and described in the SMEFT by operators with mass dimension greater than 4. At the quark level, any SMEFT operator that contributes to $0\nu\beta\beta$ decay must induce the process $dd\rightarrow uu e^- e^-$. Every such operator must therefore contain at least six fermion fields, and so have mass dimension $d \geq 9$, with contributions to the $\pi^-\rightarrow\pi^+ e^- e^-$ decay power-suppressed by a factor of $\Lambda_\mathrm{LNV}^{d - 4}$. The dimension-9 lepton-number violating operators thus contribute to the decay at leading-order (LO) in inverse powers of $\Lambda_\mathrm{LNV}$. 

There are fourteen $SU(3)_c\times U(1)_\mathrm{EM}$-invariant dimension-9 SMEFT operators which violate lepton number and may contribute to $\pi^-\rightarrow\pi^+ e^- e^-$; they can be factorized into a 4-quark operator multiplying a leptonic operator. Of these operators, four have corresponding 4-quark operators that transform as Lorentz 4-vectors, and therefore match to the $\chi\mathrm{EFT}$ operator $\pi (\partial^\mu \pi) \overline e \gamma_\mu \gamma_5 e^\mathrm{c} + \mathrm{h.c.}$, where the superscript c denotes charge conjugation and $\pi$ and $e$ represent the pion and electron fields. Integration by parts shows that pionic matrix elements of this operator are proportional to one power of the electron mass and give subleading contributions to the decay $\pi^-\rightarrow\pi^+ e^- e^-$. Of the remaining ten operators, five have corresponding 4-quark operators with positive parity and contribute to $\pi^-\rightarrow\pi^+ e^- e^-$, while the five operators containing 4-quark operators of negative parity do not contribute. Consequently, at LO the decay is described with the Lagrangian
\begin{equation}
    \mathcal L_\mathrm{SMEFT}^{0\nu\beta\beta} = \overline{e} e^\mathrm{c} \frac{G_F^2}{\Lambda_\mathrm{LNV}} \sum_k c_k \mathcal O_k,
    \label{eq:smeft_lagrangian}
\end{equation}
where $G_F$ is the Fermi coupling constant, $c_k$ are dimensionless Wilson coefficients, and the operator basis $\mathcal \{\mathcal O_k(x)\}$ is
\begin{align}
\begin{split}
	\mathcal O_1(x) &= (\overline q_L(x)\tau^+\gamma^\mu q_L(x)) [\overline q_R(x)\tau^+\gamma_\mu q_R(x)]  \\
	\mathcal O_2(x) &= (\overline q_R(x)\tau^+ q_L(x)) [\overline q_R(x)\tau^+ q_L(x)] \\
			&\hspace{0.5cm} + (\overline q_L(x)\tau^+ q_R(x)) [\overline q_L(x)\tau^+ q_R(x)] \\
	\mathcal O_3(x) &= (\overline q_L(x)\tau^+ \gamma^\mu q_L(x)) [\overline q_L(x)\tau^+ \gamma_\mu q_L(x)] \\
			&\hspace{0.5cm} + (\overline q_R(x)\tau^+ \gamma^\mu q_R(x)) [\overline q_R(x)\tau^+ \gamma_\mu q_R(x)] \label{eq:short_distance_operators} \\
	\mathcal O_{1'}(x) &= (\overline q_L(x)\tau^+\gamma^\mu q_L(x)] [\overline q_R(x)\tau^+\gamma_\mu q_R(x)) \\
	\mathcal O_{2'}(x) &= (\overline q_R(x)\tau^+ q_L(x)] [\overline q_R(x)\tau^+ q_L(x)) \\
			&\hspace{0.5cm} + (\overline q_L(x)\tau^+ q_R(x)] [\overline q_L(x)\tau^+ q_R(x)),
\end{split}
\end{align}
with $k\in \{1, 2, 3, 1', 2'\}$~\cite{PhysRevD.68.034016}. Here $q_L(x)$ and $q_R(x)$ are the left and right-handed components of the quark field isospin doublet, respectively, and
\begin{equation}
    \tau^+ = \begin{pmatrix} 0 & 1 \\ 0 & 0 \end{pmatrix}
\end{equation} 
is the isospin raising operator. The round and square brackets in Eq.~\eqref{eq:short_distance_operators} denote color contraction: for arbitrary Dirac matrices $\Gamma_1$ and $\Gamma_2$, the operators $\{\mathcal{O}_1(x), \mathcal{O}_2(x), \mathcal{O}_3(x)\}$ factor into products of color singlets, $(\overline{u}\Gamma_1 d) [\overline u \Gamma_2 d]\equiv (\overline{u}^a\Gamma_1 d^a) (\overline u^b \Gamma_2 d^b)$, whereas the operators  $\{\mathcal{O}_{1'}(x), \mathcal{O}_{2'}(x)\}$ mix color between the two Dirac bilinear terms, $(\overline{u} \Gamma_1 d][\overline u \Gamma_2 d)\equiv (\overline{u}^a\Gamma_1 d^b) (\overline u^b \Gamma_2 d^a)$, where $a, b$ are color indices. The operator basis $\{\mathcal O_k(x)\}$ of Eq.~\eqref{eq:short_distance_operators} is named the BSM basis and is typically used in phenomenological calculations of $0\nu\beta\beta$ decay~\cite{Cirigliano:2020yhp}.

Although the $\pi^-\rightarrow\pi^+ e^- e^-$ transition is unphysical, it has phenomenological importance as it can be related to the nuclear decays with $\chi\mathrm{EFT}$~\cite{Cirigliano:2018yza}. In particular, the two-nucleon decay $n^0 n^0\rightarrow p^+ p^+ e^- e^-$ is induced in $\chi$EFT by the diagrams in Fig.~(\ref{fig:0nubb_modes}) and has LO contributions from the $\pi\pi$ and $NN$ vertices~\cite{Cirigliano:2020yhp,Cirigliano:2017djv}.\footnote{Earlier work~\cite{PhysRevD.68.034016} found the $\pi\pi$ contributions to be the sole LO contribution, but the Weinberg power counting used therein did not account for regulator dependence completely.} The associated effective Lagrangian relevant for $\pi^-\rightarrow \pi^+ e^- e^-$ (i.e., omitting $NN$ and $\pi N$ operators which do not contribute) is~\cite{PhysRevLett.121.172501},
\begin{align} \begin{split}
	\mathcal L_{\chi\mathrm{EFT}}^{0\nu\beta\beta} = \overline{e} e^\mathrm{c} \frac{G_F^2}{\Lambda_\mathrm{LNV}} \frac{\Lambda_{\chi}^4}{(4\pi)^2} \frac{f_\pi^2}{8} \bigg( c_1 \beta_1 \mathcal O_1^\chi - & \frac{c_2\beta_2}{2} \mathcal O_2^\chi \\ - c_3 \beta_3 \mathcal O_3^\chi
	+ c_{1'} \beta_{1'} \mathcal O_{1'}^{\chi} - & \frac{c_{2'}\beta_{2'}}{2} \mathcal O_{2'}^{\chi} \bigg). 
	\label{eq:chipt_lagrangian}
\end{split} \end{align}
Here, $f_\pi$ is the pion decay constant in the chiral limit, $\Lambda_{\chi}^2\equiv 8\pi^2 f_\pi^2$ is the scale of chiral symmetry breaking, and $\mathcal O_k^\chi$ denote the leading $\chi\mathrm{EFT}$ operators corresponding to $\mathcal O_k$~\cite{Scherer:2002tk}. The $\chi\mathrm{EFT}$ LECs $\beta_k$ determine the $\pi\pi$ coupling, and are also essential input to study the nucleonic decay. The $\beta_k$ can be determined by evaluating the pion matrix elements of the $\mathcal O_k$ in LQCD and matching them to the corresponding matrix elements of $\mathcal O_k^\chi$ in Eq.~\eqref{eq:chipt_lagrangian}.

\subsection{Bare matrix elements}
\label{sec:bare_mat_elems}

\begin{table*}[!t]
\setlength{\tabcolsep}{5pt}
\centering
\begin{tabular}{ ccc | ccc | cc | cc }
\hline
\hline
   Ensemble & \( a m_{l} \) & \( a m_{s} \) & \( \beta \) & \( L^{3} \times T \times L_{s} \) & \( a \) [fm] & \( m_{\pi} \) [MeV] & $f_\pi$ [MeV] & $\mathcal Z_A$ & \( \mathcal Z_V \) \\
\hline 
   \multirow{2}{*}{{\vspace{-1.1mm}24I}} & 0.01 & \multirow{2}{*}{{\vspace{-1.1mm}0.04}} & \multirow{2}{*}{{\vspace{-1.1mm}2.13}} & \multirow{2}{*}{{\vspace{-1.1mm}\(24^{3} \times 64 \times 16\)}} & \multirow{2}{*}{{\vspace{-1.1mm}0.1106(3)}} & 432.2(1.4) & 163.72(64) & \multirow{2}{*}{{\vspace{-1.1mm}0.71670(22)}} & \multirow{2}{*}{{\vspace{-1.1mm}0.71273(26)}} \\
   & 0.005 & & & & & 339.6(1.2) & 151.55(62) & & \\
\hline
  \multirow{3}{*}{{\vspace{-1.7mm}32I}} & 0.008 & & &  & & 410.8(1.5) & 162.02(90) & \multirow{3}{*}{{\vspace{-1.7mm}0.74482(15)}} &  \multirow{3}{*}{{\vspace{-1.7mm}0.7440(18)}} \\
   & 0.006 & 0.03 & 2.25 & \(32^{3} \times 64 \times 16\) & 0.0828(3) & 359.7(1.2) & 154.28(70) & & \\
   & 0.004 & & & & & 302.0(1.1) & 147.54(81) & & \\
\hline
\hline
\end{tabular}
	\caption{
	Parameters of the gauge field ensembles used in this study. Each ensemble was generated with two degenerate light quark flavors of mass $m_\ell$ and one heavy quark flavor of mass $m_s$. The lattice volumes are $L^3\times T\times L_s$, with the fifth dimension having $L_s$ sites. Derived quantities are computed in Ref.~\cite{Detmold:2020jqv} (the pion mass $m_\pi$, the pion decay constant $f_\pi$, and the axial current renormalization $\mathcal Z_A$) and Refs.~\cite{PhysRevD.93.054502, PhysRevD.93.074505} (the vector current renormalization $\mathcal Z_V$ and the inverse lattice spacing $a^{-1}$).
	}
	\label{table:setup}
\end{table*}

The pion matrix elements of each of the SMEFT operators in Eq.~\eqref{eq:short_distance_operators} are computed in LQCD using gauge-field ensembles with $N_f = 2 + 1$ quark flavors generated by the RBC/UKQCD collaboration~\cite{PhysRevD.78.114509,RBC:2010qam}. Each ensemble uses the Shamir kernel~\cite{Shamir:1993zy} for the domain-wall fermion action~\cite{Kaplan:1992bt} and the Iwasaki action~\cite{IWASAKI1984449} for the gauge field. The parameters of each ensemble are detailed in Table~\ref{table:setup}, and additional details regarding the ensemble generation can be found in Refs.~\cite{PhysRevD.78.114509, RBC:2010qam, Boyle:2015exm}. The scale is set using the Wilson flow scale $w_0$~\cite{PhysRevD.93.054502}. The pion mass, $m_\pi$, the pion decay constant, $f_\pi$, and the axial-vector renormalization constant, $\mathcal{Z}_A$, for each ensemble were determined in Ref.~\cite{Detmold:2020jqv}. In the conventions used here, the physical pion decay constant~\cite{ParticleDataGroup:2020ssz} is $f_{\pi}^{(\mathrm{phys})} = 130.2\;\mathrm{MeV}$. 
The vector renormalization constant, $\mathcal Z_V$, for these ensembles was computed in the chiral limit in Refs.~\cite{PhysRevD.93.054502, PhysRevD.93.074505}. Because $\mathcal Z_V\approx\mathcal Z_A$, the ensembles exhibit approximate chiral symmetry.

On each ensemble, the time-averaged two-point function
\begin{equation}
    \mathcal C_\mathrm{2pt}(t) = \frac{1}{T}\sum_{t_- = 0}^{T - 1}\sum_{\bm x, \bm y} \langle 0 | \chi_\pi(\bm x, t + t_-) \chi_\pi^\dagger(\bm y, t_-) |0\rangle \label{eq:twopt_corr}
\end{equation}
and three-point functions
\begin{equation}
    \mathcal C_{k}(t_-, t_x, t_+) = \sum_{\bm x, \bm y, \bm z} \langle 0 | \chi_{\pi}^\dagger(\bm x, t_+) \mathcal O_k(\bm z, t_x) \chi_{\pi}^\dagger(\bm y, t_-) |0\rangle, \label{eq:threept_corr_dfn}
\end{equation}
where the pion interpolating operator $\chi_{\pi}(x) = \overline{u}(x)\gamma_5 d(x)$ has the quantum numbers of the $\pi^-$ and $t_+ \geq t_x\geq t_-$, are computed for each operator $\mathcal O_k(x)$ in the BSM basis (Eq.~\eqref{eq:short_distance_operators}). Wall-source propagators are computed at each available time slice on each configuration, where ``wall" denotes projection to vanishing three-momentum in the Coulomb gauge. Note that wall sources are not gauge-invariant, hence the need for gauge fixing. The two-point functions (Eq.~\eqref{eq:twopt_corr}) are constructed using a wall source propagator at $t_-$ and a wall sink at $t + t_-$, and the three-point functions (Eq.~\eqref{eq:threept_corr_dfn}) are constructed using wall source propagators at $t_-$ and $t_+$ and a point (local) sink at $t_x$. The explicit Wick contractions are given in Appendix~\ref{appendix:threept_contractions}.  

The bare pion matrix elements in lattice units
\begin{equation}
\langle\mathcal{O}_k\rangle\equiv a^4\langle\pi^+ | \mathcal O_k(\bm p = \bm 0) | \pi^-\rangle = a^4 \sum_{\bf x} \langle\pi^+ | \mathcal O_k(\mathbf{x}, 0) | \pi^-\rangle
\label{eq:mat_elems_latt_units}
\end{equation}
are extracted from the effective matrix elements
\begin{equation}
     O_k^{\eff}(t)\equiv 2 m_\pi \frac{\mathcal C_k(0, t, 2t)}{\mathcal C_{2\mathrm{pt}}(2t) - \frac{1}{2} \mathcal C_{2\mathrm{pt}}(T / 2) e^{m_\pi (2t - T / 2)}}.
    \label{eq:eff_mat_elem}
\end{equation}
Subtracting $\frac{1}{2}\mathcal C_{2\mathrm{pt}}(T / 2) e^{m_\pi (2t - T / 2)}$ in the denominator of Eq.~\eqref{eq:eff_mat_elem} isolates the backwards-propagating state in the two-point function, and in the $0\ll t \ll T$ limit $O_k^{\eff}(t)$ asymptotes to $\langle\mathcal{O}_k\rangle$. The effective matrix elements are computed on between 33 and 53 gauge field configurations for each ensemble (details in Appendix~\ref{appendix:r_ratio_fits}, Table~\ref{table:ratio_fits}), resampled using a bootstrap procedure with $n_b = 50$ bootstrap samples. The spectral decomposition of $O_k^\eff(t)$ up to and including the first excited state with energy $m_\pi + \Delta$,
\begin{equation}
    O_k^{\eff}(t) = \frac{\langle\mathcal{O}_k\rangle + \mathcal N_1^{(k)} e^{-\Delta t} + \mathcal N_2^{(k)} e^{-(m_\pi + \Delta)(T - 2t)}}{1 + \mathcal N_3^{(k)} e^{-2\Delta t} + \mathcal N_4^{(k)} e^{-(m_\pi + \Delta)T + 2(2m_\pi + \Delta)t}},
    \label{eq:spectral_decomp}
\end{equation}
parameterizes the ground and excited-state contributions to $O_k^{\eff}(t)$, where the coefficients $\mathcal N_i^{(k)}$ are constants determined by the spectral content of the theory. Eq.~\eqref{eq:spectral_decomp} can be Taylor expanded to first order in $\mathcal N_3^{(k)}$ and $\mathcal N_4^{(k)}$, yielding
\begin{align} \begin{split}
    f_k(t; \langle\mathcal O_k\rangle, & \, m^{(k)}, \Delta^{(k)}, A_i^{(k)}) \equiv \langle\mathcal{O}_k\rangle + A_1^{(k)} e^{-\Delta^{(k)} t} \\ 
    & + A_2^{(k)} e^{-(m^{(k)} + \Delta)(T - 2t)} - A_3^{(k)} e^{-2\Delta^{(k)} t} \\
    & - A_4^{(k)} e^{-(m^{(k)} + \Delta)T + 2(2m^{(k)} + \Delta^{(k)})t}. \label{eq:fit_model}
\end{split} \end{align}
This function is used to model the temporal dependence of $O_k^{\eff}(t)$, treating $\langle\mathcal O_k\rangle, \, m^{(k)}, \Delta^{(k)}$, and $A_i^{(k)}$ as free parameters.

\begin{figure}[!t]
    \centering
    \includegraphics[clip]{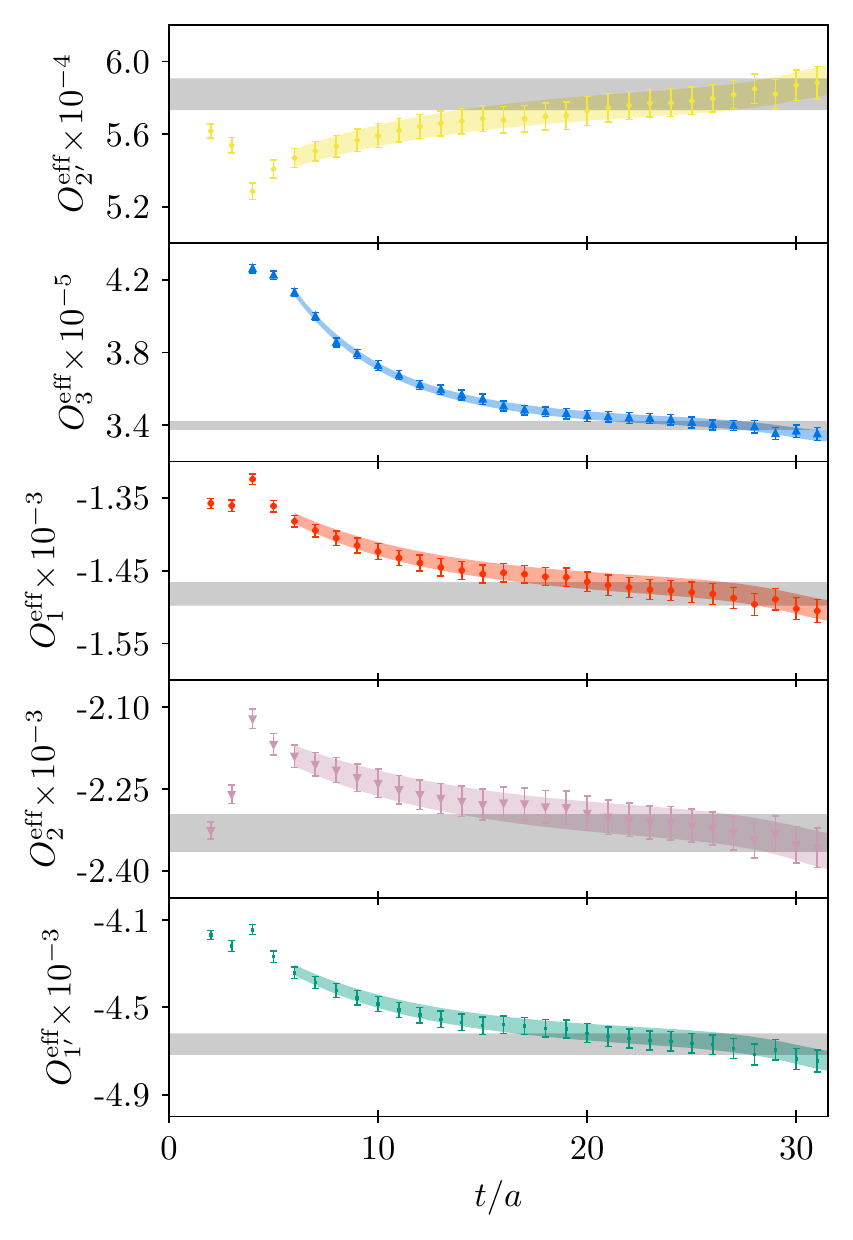}
    \caption{Effective matrix elements $O_k^\eff(t)$ (Eq.~\eqref{eq:eff_mat_elem}) computed on the 32I, $am_\ell = 0.004$ ensemble. Colored bands denote the best-fit band for the corresponding excited-state fit to the model of Eq.~\eqref{eq:fit_model}, with $[t_\mathrm{min}, t_\mathrm{max}] = [6, 32]$ and $\lambda = 0.1$. The grey band in each panel denotes the extracted value of $\langle\mathcal O_k\rangle$ (Eq.~\eqref{eq:fit_model}).
    }
    \label{fig:ex_r_ratio_fits}
\end{figure}

Fits of $O_k^{\eff}(t)$ to the model of Eq.~\eqref{eq:fit_model} are performed using a correlated least-squares fit. Each fit is performed over a given range $[t_\mathrm{min}, t_\mathrm{max}]$, with the covariance matrix 
obtained from the bootstrapped sample covariance matrix via linear shrinkage with parameter $\lambda$~\cite{Ledoit:2004,Rinaldi:2019thf}; the hyperparameters are varied, with $t_\mathrm{min}\in [6, 11]$, $t_\mathrm{max}\in [30, 32]$, and $\lambda\in \{0.1, 0.2, 0.3, 0.4\}$. Bayesian priors are placed on the model parameters, informed by the results of a two-state fit to $C_{2\mathrm{pt}}(t)$. The priors on the spectral coefficients are set to $A_k^{(i)} = 0.0\pm 0.1$, where $\mu \pm \sigma$ denotes the normal distribution with mean $\mu$ and width $\sigma$. To enforce positivity, log-normal priors are chosen for the mass $m_\pi^{(k)}$ and excited state gap $\Delta^{(k)}$ such that $m^{(k)} = m_\pi\pm \delta m_\pi$, where $m_\pi$ ($\delta m_\pi$) is the mean (standard deviation) of the pion mass (Table~\ref{table:setup}), and $\Delta^{(k)} = 2m_\pi\pm m_\pi$. Statistically indistinguishable results are obtained for $\langle\mathcal O_k\rangle$ under variation of all hyperparameters within the ranges described above, and when widths of the priors are inflated by a factor of 2, hence fiducial values of the hyperparameters are chosen as $[t_\mathrm{min}, t_\mathrm{max}] = [6, 32]$ and $\lambda = 0.1$\footnote{This choice for $\lambda$ is statistically the most conservative within the range, as $\lambda = 0$ corresponds to no shrinkage.}. Posterior values for $A_3^{(k)}$ and $A_4^{(k)}$ are found to be $\ll 1$, thus the Taylor expansion in Eq.~\eqref{eq:fit_model} is valid. The fits have $\chi^2/\mathrm{dof}$ between 0.10 and 0.73. Fit results and the complete set of fits for each operator on each ensemble with the fiducial hyperparameters are shown in Appendix~\ref{appendix:r_ratio_fits}. Illustrative fits to data from the 32I, $am_\ell = 0.004$ ensemble with the fiducial hyperparameters are shown in Fig.~\eqref{fig:ex_r_ratio_fits}. 

\subsection{Renormalization} 
\label{sec:renorm}

To make contact with phenomological calculations, lattice-regulated matrix elements must be renormalized in the $\overline{\mathrm{MS}}$ scheme. In this calculation, the renormalization coefficients are computed non-perturbatively in the RI/sMOM-$(\gamma^\mu, \gamma^\mu)$ (abbreviated as RI$\gamma$) scheme~\cite{Sturm:2009kb,Boyle:2017skn} and perturbatively matched to $\overline{\mathrm{MS}}$. In terms of the operator basis $\{\mathcal O_k(x)\}$ (Eq.~\eqref{eq:short_distance_operators}), the renormalized matrix elements can be expressed as
\begin{align} \begin{split}
    \mathcal{O}_k^{\MS}(x&; \mu^2, a) = \mathcal{Z}_{k\ell}^{\MS; \mathcal{O}}(\mu^2, a) \mathcal{O}_\ell(x; a) \\
    &= \mathcal C^{\MS\leftarrow\mathrm{RI}\gamma;\mathcal{O}}_{kj}(\mu^2, a) \mathcal{Z}_{j\ell}^{\mathrm{RI}\gamma;\mathcal{O}}(\mu^2, a) \mathcal{O}_\ell(x; a),
    \label{eq:renorm_coeff_dfn}
\end{split} \end{align}
where sums over repeated indices are implied. Here $\mathcal O_\ell(x; a)$ denotes the bare operator at lattice spacing $a$, and
\begin{equation}
    \mathcal C_{kj}^{\MS\leftarrow\mathrm{RI}\gamma;\mathcal{O}}(\mu^2, a)\equiv \mathcal Z^{\MS;\mathcal{O}}_{ki}(\mu^2, a) \left[ \mathcal Z^{\mathrm{RI}\gamma;\mathcal{O}}(\mu^2, a)\right]^{-1}_{ij}
    \label{eq:matching_coeff}
\end{equation}
is the multiplicative matching coefficient from the RI$\gamma$ to $\MS$ schemes, computed at one-loop in perturbation theory in the strong coupling $\alpha_s(\mu)$~\cite{Boyle:2017skn,PhysRevD.84.014001}. Note that each renormalization coefficient is mass-independent and defined in the chiral limit. 

The renormalization coefficients, Eq.~\eqref{eq:renorm_coeff_dfn}, are conventionally computed in the Non-Perturbative Renormalization (NPR) operator basis, $\{Q_n(x)\}$, which contains different linear combinations of operators than the BSM basis of Eq.~\eqref{eq:short_distance_operators}. Correlation functions involving the color-mixed operators $\mathcal{O}_{1'}(x), \mathcal{O}_{2'}(x)$ may be rewritten with Fierz identities~\cite{Borodulin:2017pwh} as combinations of color-unmixed quark bilinears, which simplifies the calculation. The NPR basis is defined in terms of the quark bilinears:
\begin{align}
\begin{split}
    SS(x) &= (\overline u(x) d(x))(\overline u(x) d(x)), \\ 
    PP(x) &= (\overline u(x) \gamma_5 d(x))(\overline u(x) \gamma_5 d(x)), \\
    VV(x) &= (\overline u(x) \gamma_\mu d(x))(\overline u(x) \gamma^\mu d(x)), \\ 
    AA(x) &= (\overline u(x) \gamma_\mu \gamma_5 d(x))(\overline u(x) \gamma^\mu\gamma_5 d(x)), \\
    TT(x) &= \sum_{\mu < \nu} (\overline u(x) \gamma_\mu\gamma_\nu d(x))(\overline u(x)\gamma^\mu\gamma^\nu d(x)),
\end{split}
\end{align}
as 
\begin{align}
	\begin{pmatrix} Q_1(x) \\ Q_2(x) \\ Q_3(x) \\ Q_4(x) \\ Q_5(x) \end{pmatrix} \equiv \begin{pmatrix} VV(x) + AA(x) \\ VV(x) - AA(x) \\ SS(x) - PP(x) \\ SS(x) + PP(x) \\ TT(x) \end{pmatrix}.
	\label{eq:bsm_npr_bases}
\end{align}
This basis is related to the positive-parity projection of the BSM basis, Eq.~\eqref{eq:short_distance_operators}, as
\begin{align} \begin{split}
	\begin{pmatrix} Q_1(x) \\ Q_2(x) \\ Q_3(x) \\ Q_4(x) \\ Q_5(x) \end{pmatrix}
		= \begin{pmatrix} 
		0 & 0 & 2 & 0 & 0 \\
		4 & 0 & 0 & 0 & 0 \\
		0 & 0 & 0 & -2 & 0 \\
		0 & 2 & 0 & 0 & 0 \\
		0 & 2 & 0 & 0 & 4 
	\end{pmatrix} \begin{pmatrix} \mathcal{O}_1(x) \\ \mathcal{O}_2(x) \\ \mathcal{O}_3(x) \\ \mathcal{O}_{1'}(x) \\ \mathcal{O}_{2'}(x) \end{pmatrix}. \label{eq:change_of_basis}
\end{split} \end{align}
The space spanned by $\{Q_n(x)\}$ splits into three irreducible subspaces under chiral symmetry, with bases $\{Q_1(x)\}$, $\{Q_2(x), Q_3(x)\}$, and $\{Q_4(x), Q_5(x)\}$. As both the $\MS$ and RI$\gamma$ schemes obey chiral symmetry, the renormalization coefficients $\mathcal Z_{nm}^{\MS; Q}(\mu^2; a)$ and $\mathcal Z_{nm}^{\mathrm{RI}\gamma; Q}(\mu^2; a)$, which satisfy analogous equations to Eqs.~\eqref{eq:renorm_coeff_dfn} and~\eqref{eq:matching_coeff}, each factorize into a direct sum of three block diagonal matrices, each of which spans an irreducible subspace. 

\begin{figure}[!t]
	\centering
	\begin{tikzpicture}[
	  arrowlabel/.style={
	    /tikzfeynman/momentum/.cd, 
	    arrow shorten=#1,arrow distance=1.5mm
	  },
	  arrowlabel/.default=0.5
	]
	  \begin{feynman}
	    \vertex [blob] (a1) at (0, 0);
	    \node[red, crossed dot] (a) at (0, 0) {};
	    \vertex (olabel) at (0, -0.2*\renormdiasize) {\(\mathcal{O}\)};
	    \vertex (u2) at (\renormdiasize, -\renormdiasize) {\(u_c^\gamma\)};
	    \vertex (d2) at (\renormdiasize, \renormdiasize) {\(d_d^\sigma\)};
	    \vertex (u1) at (-\renormdiasize, -\renormdiasize) {\(d_b^\beta\)};
	    \vertex (d1) at (-\renormdiasize, \renormdiasize) {\(u_a^\alpha\)};
	    \diagram* {
	    (u1) -- [anti fermion, reversed momentum = \(p_2\)] (a) -- [fermion, momentum' = \(p_2\)] (d2),
	    (d1) -- [fermion, momentum' = \(p_1\)] (a) -- [anti fermion, reversed momentum = \(p_1\)] (u2),
	    };
	  \end{feynman}
	\end{tikzpicture}
    \caption{Kinematics for operator renormalization. The red crossed circle denotes the operator $\mathcal O$ which injects momentum $2q$ into the vertex, while the solid lines denote up quarks, with momentum $p_1$ into the vertex, and down quarks, with momentum $p_2$ out of the vertex. The momenta are chosen subject to the symmetric constraint, Eq.~\eqref{eq:sym_constraint}.}
    \label{fig:operator_renorm}
\end{figure}
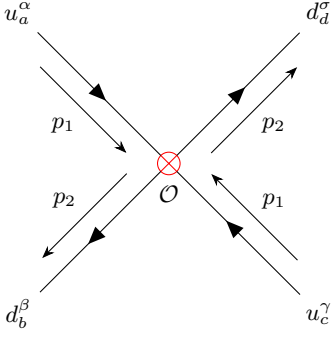

To renormalize the NPR basis operators, the four-point functions
\begin{widetext}
\begin{equation}
	(G_n)_{abcd}^{\alpha\beta\gamma\delta} (q; a, m_\ell) \equiv \frac{1}{V}\sum_{x}\sum_{x_1, ..., x_4} e^{i (p_1\cdot x_1 - p_2 \cdot x_2 + p_1\cdot x_3 - p_2\cdot x_4 + 2q\cdot x)} \langle 0 | \overline d^\delta_d (x_4) u^\gamma_c (x_3) Q_n(x) \overline d^\beta_b(x_2) u^\alpha_a(x_1) | 0\rangle
    \label{eq:op_correlator}
\end{equation}
\end{widetext}
are computed on each ensemble, where $V = L^3\times T$ is the lattice volume and $q = p_2 - p_1$. Latin letters $a, b, c, d$ denote color indices, while Greek letters $\alpha, \beta, \gamma, \delta$ denote Dirac indices. All correlation functions used for the renormalization are computed in the Landau gauge with momentum sources~\cite{Gockeler:2010yr} using 10 configurations for each ensemble, as the $V^2$ averaging from the momentum sources significantly reduces noise. The momenta are chosen subject to the symmetric constraint
\begin{equation}
    p_1^2 = p_2^2 = q^2 = \mu^2,
    \label{eq:sym_constraint}
\end{equation}
with the particular choice
\begin{align}
	p_1 = \frac{2\pi}{aL} (-j, 0, j, 0) &&
	p_2 = \frac{2\pi}{aL} (0, j, j, 0),
	\label{eq:mom_modes}
\end{align}
with $q = p_2 - p_1$ and $j\in\mathbb Z$. The kinematic configuration corresponding to $G_n(q; a, m_\ell)$ is depicted in Fig.~(\ref{fig:operator_renorm}). Note that with this choice of momentum, each value of $q$ corresponds to a unique value of $p_1$ and $p_2$, hence functions of $(p_1, p_2, q)$ are labeled as functions of $q$ for conciseness. The four-point functions are amputated,
\begin{align}
\begin{split}
    (\Lambda_n&)^{\alpha\beta\gamma\delta}_{abcd}  (q) \equiv  (S^{-1})_{aa'}^{\alpha\alpha'}(p_1) (S^{-1})_{cc'}^{\gamma\gamma'}(p_1) \\
    	& \times  (G_n)_{a'b'c'd'}^{\alpha'\beta'\gamma'\delta'}(q) (S^{-1})_{b' b}^{\beta'\beta}(p_2) (S^{-1})_{d' d}^{\delta'\delta}(p_2), 
	\label{eq:amputated_fourpt}
\end{split}
\end{align}
where
\begin{equation}
    S(p; a, m_\ell) = \frac{1}{V}\sum_{x,y} e^{ip\cdot (x - y)}\langle 0| q(x) \overline{q}(y) |0\rangle \label{eq:quark_prop}
\end{equation}
is the Landau-gauge momentum-projected quark propagator. The ensemble dependence of $\Lambda_n(q)$, $G_n(q)$, and $S(p)$ has been suppressed in Eq.~\eqref{eq:amputated_fourpt} for clarity. Projectors $(P_n)_{badc}^{\beta\alpha\delta\gamma}$ are introduced to project $(\Lambda_m)_{abcd}^{\alpha\beta\gamma\delta}$ onto the NPR basis for RI$\gamma$~\cite{Boyle:2017skn} to yield a matrix of projected four-point functions with components
\begin{equation}
    F_{mn}(q; a, m_\ell)\equiv (P_n)_{badc}^{\beta\alpha\delta\gamma} (\Lambda_m)_{abcd}^{\alpha\beta\gamma\delta}(q; a, m_\ell). \label{eq:Fmn_dfn}
\end{equation}

The remaining quantities which are computed non-perturbatively on each ensemble are the RI$\gamma$ quark-field renormalization 
\begin{equation}
    \left(\frac{\mathcal Z_q^{\mathrm{RI}\gamma}}{\mathcal Z_V}\right)(\mu^2; a, m_\ell)\bigg|_{q^2 = \mu^2} = \frac{1}{48}\mathrm{Tr}[\gamma_\mu \Lambda_V^\mu(q)], \label{eq:quark_renorm}
\end{equation}
and the vector and axial-vector-renormalization coefficients, $\mathcal Z_V(\mu^2; a, m_\ell)$ and $\mathcal Z_A(\mu^2; a, m_\ell)$, whose computation is described in Appendix~\ref{appendix:vector_axial_rcs}.
Here $\Lambda_V^\mu(q) = S^{-1}(p_1) G_V^\mu(q) S^{-1}(p_2)$ is the amputated vector three-point function, where
\begin{align} \begin{split}
    G_V^\mu(q; a, m_\ell) = \frac{1}{V}\sum_{x, x_1, x_2} & e^{i (p_1\cdot x_1 - p_2\cdot x_2  + q\cdot x)} \\
    &\times \langle 0 | u(x_1) V^\mu(x) \overline d(x_2) |0\rangle
    \label{eq:vector_correlator}
\end{split} \end{align}
is the vector three-point function, with $V^\mu(x) = \overline{u}(x) \gamma^\mu d(x)$ the vector-current. The quantities $Z\in \{\mathcal Z_q^{\mathrm{RI}\gamma} / \mathcal Z_V, F_{nm}\}$ display mild dependence on quark mass, and are extrapolated to the chiral limit via a joint fit over ensembles with different masses to the model
\begin{equation}
    Z(\mu^2; a, m_\ell) = Z(\mu^2; a) + \tilde Z(\mu^2; a) m_\ell
    \label{eq:amell_extrap_model}
\end{equation}
where $Z(\mu^2; a)$ and $\tilde Z(\mu^2; a)$ are fit coefficients, and $Z(\mu^2; a)$ is understood as the chiral limit of $Z(\mu^2; a, m_\ell)$. Correlations between $\mathcal Z_q^{\mathrm{RI}\gamma} / \mathcal Z_V$ and $F_{nm}$ on each ensemble are retained in the fits, and the covariance matrix is block-diagonal as data from different ensembles is uncorrelated. Fitted values of $Z(\mu^2; a)$ are statistically consistent when a constant model $Z(\mu^2; a, m_\ell) = Z(\mu^2; a)$ is used in place of the linear model of Eq.~\eqref{eq:amell_extrap_model}. The full set of extrapolations for $(\mathcal Z_q^{\mathrm{RI}\gamma} / \mathcal Z_V)(\mu^2; a)$ and $F_{mn}(q; a)$ for both the $a = 0.11\;\mathrm{fm}$ and $a = 0.08\;\mathrm{fm}$ ensembles is shown in Appendix~\ref{app:amell_extrapolation}.

With the definitions above, the NPR-basis renormalization coefficients in the RI$\gamma$ scheme can be computed as
\begin{equation}
    \frac{\mathcal Z_{nm}^{\mathrm{RI}\gamma; Q}}{\mathcal Z_V^2}(\mu^2; a) \bigg|_\mathrm{sym} = \left( \frac{\mathcal Z_q^{\mathrm{RI}\gamma}(\mu^2; a)}{\mathcal Z_V}\right)^2\, \left[ F_{nr}^{(\mathrm{tree})} F^{-1}_{rm}(q; a)\right]
    \label{eq:ri_smom_rc}
\end{equation}
where $F_{nr}^{(\mathrm{tree})}\equiv P_r \Lambda_n^{(\mathrm{tree})}$ is the matrix of projections of the tree-level vertex function $\Lambda_n^{(\mathrm{tree})}$, and the notation $|_\mathrm{sym}$ denotes evaluation at the symmetric kinematic point, Eq.~\eqref{eq:sym_constraint}. The renormalization coefficients $\mathcal Z_{nm}^{\mathrm{RI}\gamma; Q}(\mu^2; a)/ \mathcal Z_V^2$ are only computed non-perturbatively at scales $\mu_j = \frac{2\pi}{aL} ||(j, j, 0, 0)||$ corresponding to the lattice momenta given in Eq.~\eqref{eq:mom_modes}, where $||\cdot||$ denotes the Euclidean norm of the lattice vector. However, the matching coefficients $\mathcal C_{nm}^{\MS\leftarrow\mathrm{RI}\gamma; Q}(\mu^2, a)$ in Eq.~\eqref{eq:renorm_coeff_dfn} have been computed at $\mu = M\equiv 3\;\mathrm{GeV}$~\cite{Boyle:2017skn,PhysRevD.84.014001}, and therefore the renormalization coefficients must be perturbatively evolved from $\mu_j$ to $M$. To minimize the artifacts from truncating the perturbative expansion of the matching coefficients, $\mu_j$ must be chosen to lie in the Rome-Southampton window~\cite{Arthur:2010ht, PhysRevD.85.014501},
\begin{equation}
	\Lambda_\mathrm{QCD} \ll \mu_j \ll \left(\frac{\pi}{a}\right),
\end{equation}
with $\mu_j$ taken to satisfy $\mu_j\leq M$ to minimize discretization artifacts. In practice, the scale $\mu_4$ is used for renormalization at both $a = 0.11\;\mathrm{fm}$ and $a = 0.08\;\mathrm{fm}$, as this is the nearest available scale to $M$ satisfying these constraints. Numerically, these scales are $\mu_4 =  2.64\, \mathrm{GeV}$ for the $a = 0.11\;\mathrm{fm}$ ensemble and $\mu_4 = 2.65\;\mathrm{GeV}$ for the $a = 0.08\;\mathrm{fm}$ ensemble. Scale evolution from $\mu_4$ to $M$ is performed by integrating the evolution equation,
\begin{align} \begin{split}
	\left(\frac{\mathcal Z_{nm}^{\mathrm{RI}\gamma; Q} }{\mathcal Z_V^2} \right) & (M; a) = \left(\frac{\mathcal Z_{nm}^{\mathrm{RI}\gamma; Q} }{\mathcal Z_V^2} \right)(\mu_4; a) \\ 
	+ & \int_{\mu_4}^{M} \frac{d\mu}{\mu} \gamma_{np}^{\mathrm{RI}\gamma; Q} (\alpha_s(\mu)) \left( \frac{\mathcal Z_{pm}^{\mathrm{RI}\gamma; Q} }{\mathcal Z_V^2}(\mu; a) \right), \label{eq:scale_evolution}
\end{split} \end{align}
where the NPR basis anomalous dimensions $\gamma_{nm}^{\mathrm{RI}\gamma; Q} (\alpha_s(\mu))$ have been computed at two-loop order in $\alpha_s(\mu)$ in Ref.~\cite{Papinutto:2016xpq}. Statistically consistent results for $(\mathcal Z_{nm}^{\mathrm{RI}\gamma; Q} / \mathcal Z_V^2 ) (M)$ are obtained when $\mu_3$ is instead used as the non-perturbative scale in Eq.~\eqref{eq:scale_evolution}. 

The results for the NPR basis renormalization coefficients, computed at $\mu = 3\;\mathrm{GeV}$ in $\overline{\mathrm{MS}}$, are
\begin{widetext} \begin{align} \begin{split}
    \left(\frac{\mathcal{Z}^{\MS; Q}}{\mathcal{Z}_V^2}\right)\left(\mu^2 = 9\;\mathrm{GeV}^2, a = 0.11\;\mathrm{fm}\right) &= \begin{pmatrix}
        0.90746(43) & 0 & 0 & 0 & 0 \\
        0 & 1.04052(14) & 0.26154(56) & 0 & 0 \\
        0 & 0.05286(12) & 0.95333(75) & 0 & 0 \\
        0 & 0 & 0 & 0.91775(71) & -0.02367(13) \\
        0 & 0 & 0 & -0.28140(66) & 1.13952(35)
    \end{pmatrix}, \\
    \left(\frac{\mathcal{Z}^{\MS; Q}}{\mathcal{Z}_V^2}\right)\left(\mu^2 = 9\;\mathrm{GeV}^2, a = 0.08\;\mathrm{fm}\right) &= \begin{pmatrix}
        0.92625(51) & 0 & 0 & 0 & 0 \\
        0 & 1.03941(31) & 0.27661(50) & 0 & 0 \\
        0 & 0.04203(67) & 0.85916(82) & 0 & 0 \\
        0 & 0 & 0 & 0.84035(87) & -0.01061(40) \\
        0 & 0 & 0 & -0.29928(71) & 1.19362(57)
    \end{pmatrix}. \label{eq:renorm_coeffs_msbar}
\end{split} \end{align} \end{widetext}
The components corresponding to transitions between operators in different irreducible chiral representations are consistent with $|\mathcal Z_{nm}^{\MS; Q} / \mathcal{Z}_V^2| < 10^{-5}$ and thus set to zero in Eq.~\eqref{eq:renorm_coeffs_msbar}. The renormalization coefficients have been computed for the NPR operator basis (Eq.~\eqref{eq:bsm_npr_bases}) in Ref.~\cite{Boyle:2017skn} using $s$ quarks in place of $d$ quarks. The results in Ref.~\cite{Boyle:2017skn} agree with Eq.~\eqref{eq:renorm_coeffs_msbar} at the percent level, and deviations between the results are likely due to perturbative truncation errors, as Ref.~\cite{Boyle:2017skn} used non-pertubative step-scaling~\cite{Arthur:2010ht,PhysRevD.85.014501}. The NPR basis renormalization coefficients are converted to the BSM basis using the change of basis matrix, Eq.~\eqref{eq:change_of_basis}, and combined with the bare matrix elements to form renormalized matrix elements,
\begin{equation}
    \vspace{-0.005cm}
    O_k(m_\pi, f_\pi, a, L)\equiv \langle\pi^+ | \mathcal O_k^{\overline{\mathrm{MS}}}(\bm p = \bm 0) | \pi^- \rangle (m_\pi, f_\pi, a, L).
    \label{eq:renorm_mat_elems_ens}
\end{equation} 
On a given ensemble, the renormalization coefficients and bare matrix elements are computed on different configurations, as the former are only computed on a subset of 10 of the configurations used to compute the matrix elements on each ensemble. As such, they are combined as an uncorrelated product and their errors are added in quadrature. The renormalized matrix elements are shown in Table~\ref{table:renorm_matrix_elems}. 

\section{Chiral extrapolation}
\label{sec:chiral_extrap}

\begin{table*}[!t]
    \setlength{\tabcolsep}{5pt}
    \centering
    \begin{tabular}{ cc | ccccc } 
    \hline\hline
    \multicolumn{2}{c|}{Operator} & $\mathcal O_1$ & $\mathcal O_2$ & $\mathcal O_3$ & $\mathcal O_{1'}$ & $\mathcal O_{2'}$ \\ 
    \hline \hline 
    \rule{0cm}{0.4cm}Ensemble & $a m_\ell$ & 
    \multicolumn{5}{|c}{$O_k(m_\pi, f_\pi, a, L)$} \\ 
    \hline \rule{0cm}{0.4cm}\multirow{2}{*}{24I} & 0.01 & -0.0190(11) & -0.0467(15) & 0.001602(59) & -0.0850(32) & 0.01556(50) \\
    & 0.005 & -0.0162(11) & -0.0391(15) & 0.000815(28) & -0.0733(32) & 0.01305(45) \\
    \hline \rule{0cm}{0.4cm}\multirow{3}{*}{32I} & 0.008 & -0.0204(15) & -0.0436(18) & 0.001383(57) & -0.0863(39) & 0.01393(66) \\
    & 0.006 & -0.0179(13) & -0.0387(14) & 0.000937(39) & -0.0771(36) & 0.01239(50) \\
    & 0.004 & -0.0160(15) & -0.0347(16) & 0.000569(24) & -0.0696(37) & 0.01115(60) \\ 
    \hline \hline
    & & \multicolumn{5}{|c}{Extrapolated $O_k(m_\pi^{(\mathrm{phys})}, f_\pi^{(\mathrm{phys})}, 0, \infty )$} \\
    \hline 
    \multicolumn{2}{c|}{$\langle \pi^+ | \mathcal{O}_k^{\overline{\mathrm{MS}}} | \pi^- \rangle\;(\mathrm{GeV}^4)$} & -0.0127(16) & -0.0245(22) & 0.0000869(80) & -0.0535(48) & 0.00757(75)
    \\ \hline
    \multicolumn{2}{c|}{$\beta_k$} & -1.21(17) & -2.37(23) & 0.606(66) & -5.17(51) & 0.735(80)
    \\ 
    \multicolumn{2}{c|}{$\alpha_k$ ($\mathrm{fm}^{-2}$)} & -0.27(31) & 0.33(23) & 0.13(22) & -0.04(23) & 0.58(26)
    \\ 
    \multicolumn{2}{c|}{$c_k$} & -0.6(1.4) & -1.17(98) & 8.6(1.4) & -1.18(98) & -1.5(1.0)
    \\ 
    \multicolumn{2}{c|}{$\chi^2 / \mathrm{dof}$} & 0.02 & 0.03 & 0.22 & 0.08 & 0.03 \\
    \hline\hline
    \end{tabular} 
    \caption{Renormalized pion matrix elements
    $O_k(m_\pi, f_\pi, a, L)$, Eq.~\eqref{eq:renorm_mat_elems_ens}, of each operator $\mathcal O_k$ in the BSM basis computed on each of the ensembles (upper), and the results of chiral continuum extrapolation (lower). The parameters $\alpha_k$, $\beta_k$, and $c_k$ are the $\chi\mathrm{EFT}$ LECs, Eq.~\eqref{eq:chiral_expansion}, and $\langle \pi^+ | \mathcal{O}_k^{\overline{\mathrm{MS}}} | \pi^- \rangle$ is the extrapolated matrix element in the continuum and infinite volume limit at physical quark masses in the $\overline{\mathrm{MS}}$ scheme at $\mu = 3\;\mathrm{GeV}$.}
    \label{table:renorm_matrix_elems}
\end{table*}


\begin{figure*}[!t]
    \centering
    \includegraphics{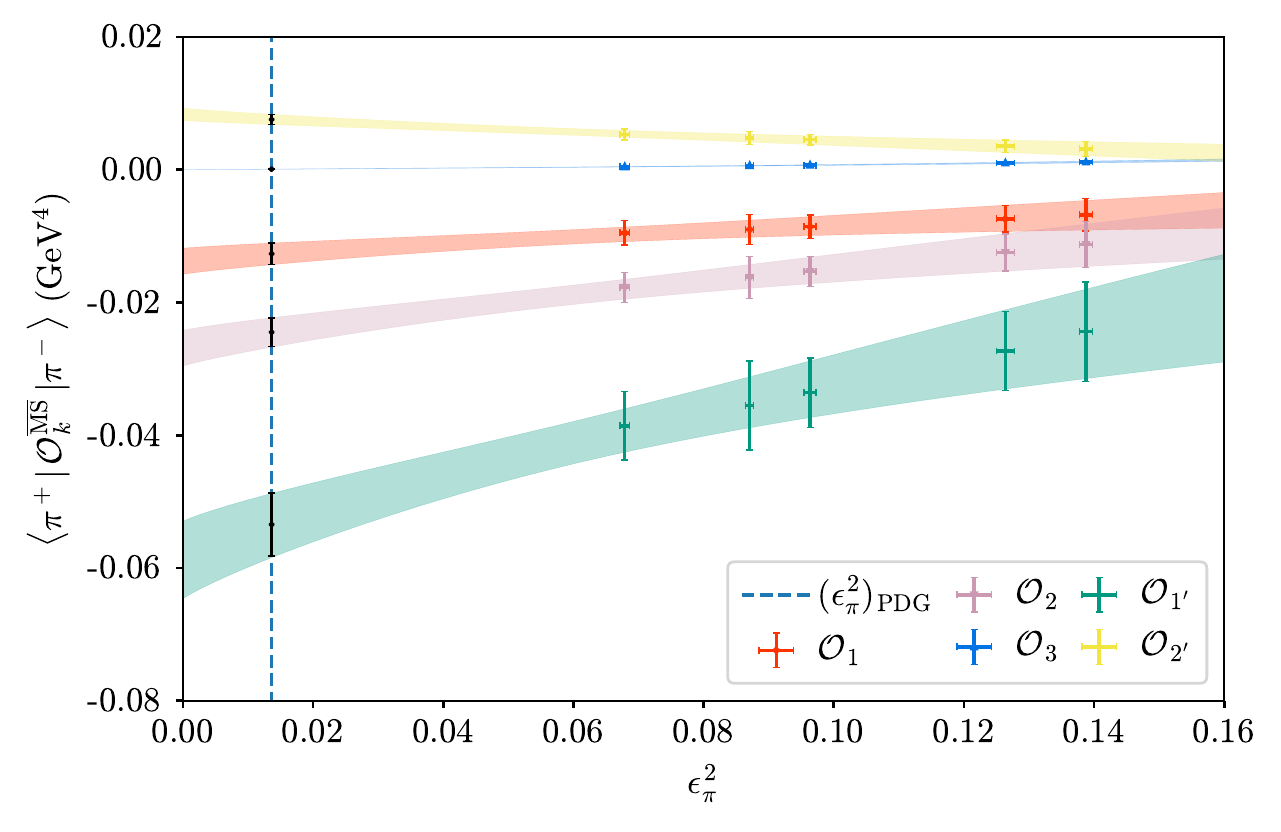}
    \caption{Chiral extrapolation of renormalized matrix elements. The LQCD results are shown at $\epsilon_\pi^2 = m_\pi^2 / (8\pi^2 f_\pi^2)$ calculated using the pion mass of each ensemble and the physical value of $f_\pi$, and the values of $O_k(m_\pi, f_\pi, a, L)$ have been shifted by $-\mathcal F_k(m_\pi, f_\pi, a, L; \alpha_k, \beta_k, c_k) + \mathcal F_k(m_\pi, f_\pi^{(\mathrm{phys})}, 0, \infty; \alpha_k, \beta_k, c_k)$, where $\alpha_k, \beta_k, c_k$ are the best-fit coefficients given in Table~\ref{table:renorm_matrix_elems}. The physical pion mass is denoted by the dashed line.}
    \label{fig:chiral_extrap}
\end{figure*}

The renormalized matrix elements $O_k(m_\pi, f_\pi, a, L)$, Eq.~\eqref{eq:renorm_mat_elems_ens}, computed on each ensemble, are extrapolated to the continuum and infinite volume limit and physical pion mass using $\chi$EFT at $\mathrm{N}^2\mathrm{LO}$; the relevant expressions have been derived in Ref.~\cite{PhysRevLett.121.172501} using the Lagrangian in Eq.~\eqref{eq:chipt_lagrangian}. The chiral models $\mathcal{F}_k$ for $O_k$ are given by
\begin{widetext}
\begin{align}
\begin{split}
    \mathcal{F}_1(m_\pi, f_\pi, a, L; \alpha_1, \beta_1, c_1) &= \frac{\beta_1\Lambda_\chi^4}{(4\pi)^2} \bigg[1 + \epsilon_\pi^2 (\log \epsilon_\pi^2 - 1
    	 + c_1 -  f_0(m_\pi L) + 2 f_1 (m_\pi L)) + \alpha_1 a^2\bigg], \\
    \mathcal{F}_2(m_\pi, f_\pi, a, L; \alpha_2, \beta_2, c_2) &= \frac{\beta_2\Lambda_\chi^4}{(4\pi)^2} \bigg[1 + \epsilon_\pi^2 (\log \epsilon_\pi^2 - 1
     + c_2 - f_0(m_\pi L) + 2 f_1 (m_\pi L)) + \alpha_2 a^2\bigg],  \\
     \mathcal{F}_3(m_\pi, f_\pi, a, L; \alpha_3, \beta_3, c_3) &= \epsilon_\pi^2 \frac{\beta_3\Lambda_\chi^4}{(4\pi)^2} \bigg[1 - \epsilon_\pi^2 (3 \log \epsilon_\pi^2 + 1 - c_3 + f_0(m_\pi L)  + 2 f_1 (m_\pi L)) + \alpha_3 a^2\bigg],
     \label{eq:chiral_expansion}
\end{split}
\end{align}
\end{widetext}
where $\epsilon_\pi^2 = m_\pi^2 / \Lambda_\chi^2$ is a power-counting parameter for $\chi\mathrm{EFT}$, $\beta_k$ are the LO LECs defined in Eq.~\eqref{eq:chipt_lagrangian}, and $\alpha_k$ and $c_k$ are the additional NLO LECs. The matrix elements $O_{1'}$ and $O_{2'}$ have the same chiral behavior as $O_1$ and $O_2$ and are modeled by $\mathcal F_1$ and $\mathcal F_2$, respectively, but with different LECs, $\alpha_{1'}, \beta_{1'}, c_{1'}$ and $\alpha_{2'}, \beta_{2'}, c_{2'}$. The functions 
\begin{align}
\begin{split}
    f_0(m L) &= -2\sum_{|\bm n|\neq 0} K_0(mL|\bm n|), \\ 
    f_1(m L) &= 4 \sum_{|\bm n|\neq 0} \frac{K_1(mL|\bm n|)}{mL|\bm n|},
\end{split}
\end{align}
are sums of modified Bessel functions $K_i(z)$ arising from one-loop, finite volume $\chi\mathrm{EFT}$ in the $p$-regime.

The models are fit to the data in Table~\ref{table:renorm_matrix_elems}, using least-squares minimization including the correlations between $O_k$, $m_\pi$, and $f_\pi$ on each ensemble. The final extrapolated results for the matrix elements and corresponding LECs are given in Table~\ref{table:renorm_matrix_elems}. The resulting fits are shown in Fig.~\eqref{fig:chiral_extrap}, where to isolate the pion-mass dependence of the matrix elements, $\epsilon_\pi^2$ has been rescaled by $(f^\mathrm{(lat)}_{\pi} / f^{(\mathrm{phys})}_\pi)^2$ and the values of $O_k(m_\pi, f_\pi, a, L)$ have been shifted by $-\mathcal F_k(m_\pi, f_\pi, a, L; \alpha_k, \beta_k, c_k) + \mathcal F_k(m_\pi, f_\pi^{(\mathrm{phys})}, 0, \infty; \alpha_k, \beta_k, c_k)$, where $\alpha_k, \beta_k, c_k$ are the best-fit coefficients given in Table~\ref{table:renorm_matrix_elems}. The extrapolation bands for each $\mathcal O_k$ depict the functional form $\mathcal F_k(m_\pi, f_\pi^{(\mathrm{phys})}, 0, \infty; \alpha_k, \beta_k, c_k)$. The results for $\langle\pi^+|\mathcal O_k^{\overline{\mathrm{MS}}} |\pi^-\rangle$ obey the same hierarchy as the chiral $SU(3)$ estimates~\cite{Cirigliano:2017ymo}, and are consistent with these results within two standard deviations.

The results for the renormalized, extrapolated, matrix elements are found to be in mild tension with the results of Ref.~\cite{PhysRevLett.121.172501}. There are a number of differences between the two calculations which may account for the discrepancy. The present calculation was performed with the same domain-wall action for the valence and sea-quarks and is thus unitary, while that of Ref.~\cite{PhysRevLett.121.172501} used a mixed action where unitarity is only restored in the continuum limit. Using the domain-wall action for valence and sea-quarks yields matrix elements that have a mild dependence on the lattice spacing. In contrast, the mixed action results appear to have a larger dependence on the lattice spacing. However, the analysis of Ref.~\cite{PhysRevLett.121.172501} was performed on nine ensembles with pion masses $m_\pi\lesssim 310\;\mathrm{MeV}$, including one ensemble with pion mass below the physical point, which allows for an interpolation to the physical point. Ref.~\cite{PhysRevLett.121.172501} also uses three lattice spacings as opposed to the two used in this computation, which allows for higher control of discretization artifacts in the non-perturbative renormalization and the chiral and continuum extrapolation.

\section{Conclusion}
\label{sec:conclusion}

This work presents a determination of the renormalized matrix elements and $\chi$EFT LECs for the short-distance operators that potentially arise from BSM physics at high scales and are relevant for the $\pi^-\rightarrow\pi^+ e^- e^-$ transition. The present calculation is the first to use chiral fermions with the same valence and sea-quark actions. The domain-wall action yields a simple renormalization coefficient structure and straightforward extrapolation to the continuum and infinite volume limit and physical value of the light quark mass. With the results of Ref.~\cite{Detmold:2020jqv}, this completes the calculation of both the long and short-distance amplitudes for $\pi^-\rightarrow\pi^+ e^- e^-$ on the same gauge-field ensembles.



One may compare the relative size of the decay amplitude of $\pi^-\rightarrow \pi^+ e^- e^-$ induced by short-distance mechanisms, $\mathcal A_\mathrm{SD}$, to that induced by long-distance mechanisms, $\mathcal A_\mathrm{LD}$. In any model with a seesaw-type mechanism~\cite{Bilenky:2012qi}, for example the minimal left-right symmetric model~\cite{Cirigliano:2018yza}, the effective Majorana neutrino mass $m_{\beta\beta}$ scales as $c / (G_F \Lambda_\mathrm{LNV})$, where $c$ is a Wilson coefficient. This implies
\begin{align} \begin{split}
    \frac{\mathcal A_\mathrm{SD}}{\mathcal A_\mathrm{LD}} &= \frac{\frac{G_F^2}{\Lambda_\mathrm{LNV}} |\sum_k c_k \langle \pi^+ | \mathcal O_k | \pi^- \rangle | }{ G_F^2 m_{\beta\beta} |M^{0\nu}|} \\
    &= G_F \frac{|\sum_k c_k \langle \pi^+ | \mathcal O_k | \pi^- \rangle |}{ c \, |M^{0\nu}|} \\
    &\sim G_F \frac{\Lambda_\mathrm{QCD}^4}{\Lambda_\mathrm{QCD}^2}\sim 10^{-5},
    \label{eq:sd_ld_ratio}
\end{split} \end{align}
where $M^{0\nu}$ is the long-distance nuclear matrix element for $\pi^-\rightarrow\pi^+ e^- e^-$. The final line of Eq.~\eqref{eq:sd_ld_ratio} arises by assuming that in a given BSM model, the dimensionless Wilson coefficients, $c_k$ and $c$, describing each amplitude are order 1, and by using dimensional arguments to approximate the matrix elements. In particular, the long-distance nuclear matrix element includes the convolution of a massless bosonic propagator with a bilocal QCD matrix element. The convolution picks out the dimensional scale $1 / \Lambda_\mathrm{QCD}^2$, thereby enhancing the long-distance contribution compared to the short-distance one. 

Since $\langle \pi^+ | \mathcal O_k | \pi^- \rangle$ and $M^{0\nu}$ have now been computed consistently in LQCD, it is possible to compute the ratio of Eq.~\eqref{eq:sd_ld_ratio}, quantitatively, given the Wilson coefficients $c_k$ and $c$ from some model. For example, taking $c_k = c = 1$, and using the LQCD results from this work and of Ref.~\cite{Detmold:2020jqv} for the matrix elements yields $\frac{\mathcal A_\mathrm{SD}}{\mathcal A_\mathrm{LD}} = 6.1(2)\times 10^{-5}$, consistent with expectations.

In addition to the pion-pion $\chi\mathrm{EFT}$ LECs, the other LECs contributing to nuclear $0\nu\beta\beta$ decay must be determined in future calculations in order to constrain models of new physics from experimental constraints on nuclear $0\nu\beta\beta$ decay rates. Knowledge of these LECs may be used as input for models of nuclear many-body physics, which may be used to estimate the half-lives of various nuclear $0\nu\beta\beta$ decay processes from short-distance mechanisms with increasing precision. The other LO LECs that are necessary for describing nuclear $0\nu\beta\beta$ decay are from the nucleon-nucleon interaction (Fig.~\eqref{fig:nn_vertex}), and may be determined with knowledge of the $\langle p^+ p^+ | \mathcal O_k(\bm p = \bm 0) | n^0 n^0\rangle$ matrix elements~\cite{Cirigliano:2020yhp}. Calculations of these matrix elements are ongoing and will provide the first direct LQCD probe of $0\nu\beta\beta$ decay in nuclear systems. 

\begin{acknowledgements}

The authors thank Zhenghao Fu and Anthony Grebe for helpful discussions, Nicolas Garron for insightful discourse regarding the renormalization, and Evan Berkowitz, Nicolas Garron, Henry Monge-Camacho, Amy Nicholson, and Andr\'e Walker-Loud for careful reading and comments on an earlier version of this manuscript. The calculation of propagators and correlation functions described in the text was performed on the IBM Blue Gene/Q computers of the RIKEN-BNL Research Center and Brookhaven National Laboratory, and on the facilities of the USQCD collaboration. These computations used the CPS~\cite{Jung:2014ata}, GLU~\cite{glu}, Grid~\cite{Boyle:2015tjk}, Chroma~\cite{Edwards:2004sx}, QLUA~\cite{qlua}, and QUDA~\cite{Clark:2009wm} software packages. Least-squares fits were performed with the lsqfit software package~\cite{peter_lepage_2021_5777652}. Furthermore, the Tikz-Feynman package~\cite{Ellis:2016jkw} was used to generate diagrams for this manuscript, and the RunDec3 package~\cite{Herren:2017osy} was used to run the strong coupling in the perturbative matching for the renormalization. 
This work is supported in part by the U.S.~Department of Energy, Office of Science, Office of Nuclear Physics under grant Contract Number DE-SC0011090. WD is also supported by the SciDAC5 award DE-SC0023116. PES is additionally supported by the National Science Foundation under EAGER grant 2035015, and by the U.S. DOE Early Career Award DE-SC0021006.

\end{acknowledgements}

\newpage

\bibliography{0nubb_writeup}

\begin{appendices}

\clearpage
\onecolumngrid
\section{Three-point contractions}
\label{appendix:threept_contractions}

The correlation functions of Eq.~\eqref{eq:threept_corr_dfn} can be written in terms of the following contraction structures,
\begin{align} \begin{split}
    \circled{1}_{\Gamma_1\Gamma_2} &= \sum_{\bm{x}} \Tr \left[ \gamma_5 \Gamma_1 S_d(t_-\rightarrow x) S_u^\dagger (t_-\rightarrow x) \right]\cdot  \Tr \left[ \gamma_5 \Gamma_2 S_d\left(t_+\rightarrow x \right) S_u^\dagger (t_+\rightarrow x)\right] + (t_-\leftrightarrow t_+), \\
    \circled{2}_{\Gamma_1\Gamma_2} &= \sum_{\bm x} \Tr [\gamma_5 \Gamma_1 S_d(t_-\rightarrow x) S_u^\dagger (t_-\rightarrow x) \gamma_5 \Gamma_2  S_d (t_+\rightarrow x) S_u^\dagger (t_+\rightarrow x)] + (t_-\leftrightarrow t_+), \\
    \circled{3}_{\Gamma_1\Gamma_2} &= \sum_{\bm x} \TrC [\TrD [\gamma_5 \Gamma_1 S_d(t_-\rightarrow x) S_u^\dagger (t_-\rightarrow x)] \cdot \TrD [\gamma_5 \Gamma_2 S_d(t_+\rightarrow x) S_u^\dagger (t_+\rightarrow x)]] + (t_-\leftrightarrow t_+), \\
    \circled{4}_{\Gamma_1\Gamma_2} &= \sum_{\bm x} \TrD [\TrC [\gamma_5 \Gamma_1 S_d(t_-\rightarrow x) S_u^\dagger (t_-\rightarrow x)] \cdot \TrC [\gamma_5 \Gamma_2 S_d(t_+\rightarrow x) S_u^\dagger (t_+\rightarrow x)]] + (t_-\leftrightarrow t_+), \label{eq:threept_contraction_structures}
\end{split} \end{align}
where $\Gamma_1, \Gamma_2$ are arbitrary Dirac matrices, $\mathrm{Tr}_\mathrm{C}$ ($\mathrm{Tr}_\mathrm{D}$) denotes a color (spin) trace, $\mathrm{Tr} = \mathrm{Tr}_\mathrm{C}\circ\mathrm{Tr}_\mathrm{D}$ denotes a full trace, and $x = (\bm x, t_x)$. Propagators $S(t_{\mathrm{src}}\rightarrow x)$ are computed with a zero three-momentum wall source at time $t_\mathrm{src}\in \{t_-, t_+\}$ and a point sink at time $t_x$,
\begin{equation}
    S(t_{\mathrm{src}}\rightarrow x)\equiv \sum_{\bm y} S((\bm y, t_{\mathrm{src}}) \rightarrow (\bm x, t_x)).
\end{equation}
With the definitions of Eq.~\eqref{eq:threept_contraction_structures}, the correlation functions are evaluated as
\begin{align}\begin{split}
    \mathcal C_{1}(t_-, t_x, t_+) &= -\frac{1}{4}\left[\circled{1}_{VV} - \circled{2}_{VV} - \circled{1}_{AV} + \circled{2}_{AV} + \circled{1}_{VA} - \circled{2}_{VA} - \circled{1}_{AA} + \circled{2}_{AA}\right], \\
    \mathcal C_{2}(t_-, t_x, t_+) &= -\frac{1}{2}\left[ \circled{1}_{SS} - \circled{2}_{SS} + \circled{1}_{PP} - \circled{2}_{PP} \right], \\
    \mathcal C_{3}(t_-, t_x, t_+) &= -\frac{1}{2}\left[ \circled{1}_{VV} - \circled{2}_{VV} + \circled{1}_{AA} - \circled{2}_{AA} \right], \\
    \mathcal C_{1'}(t_-, t_x, t_+) &= -\frac{1}{4}\left[ \circled{3}_{VV} - \circled{4}_{VV} - \circled{3}_{AV} + \circled{4}_{AV} + \circled{3}_{VA} - \circled{4}_{VA} - \circled{3}_{AA} + \circled{4}_{AA} \right], \\
    \mathcal C_{2'}(t_-, t_x, t_+) &= -\frac{1}{2}\left[ \circled{3}_{SS} - \circled{4}_{SS} + \circled{3}_{PP} - \circled{4}_{PP} \right], \label{eq:threept_contractions}
\end{split}\end{align}
where $S = 1$, $P = \gamma_5$, $V = \gamma^\mu$, and $A = \gamma^\mu\gamma_5$.

\clearpage
\section{Effective matrix element fits}
\label{appendix:r_ratio_fits}

\begin{figure*}[!htp]
    \centering
    \subfloat{\includegraphics{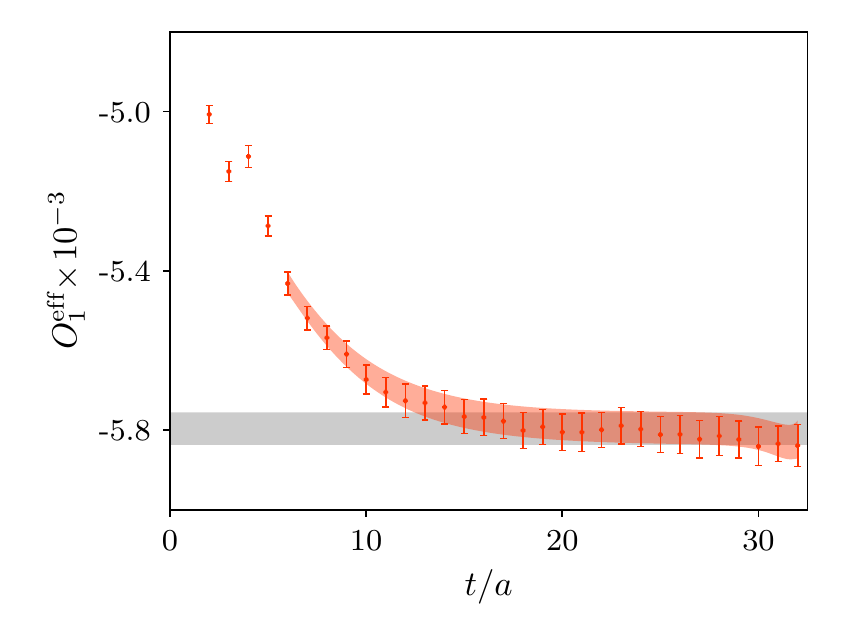}}
    \subfloat{\includegraphics{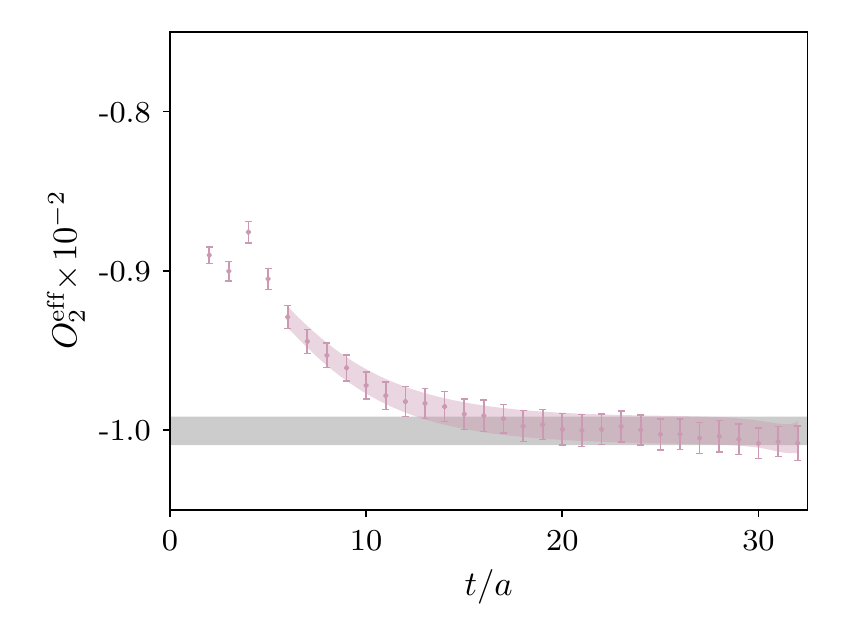}} \\
    \subfloat{\includegraphics{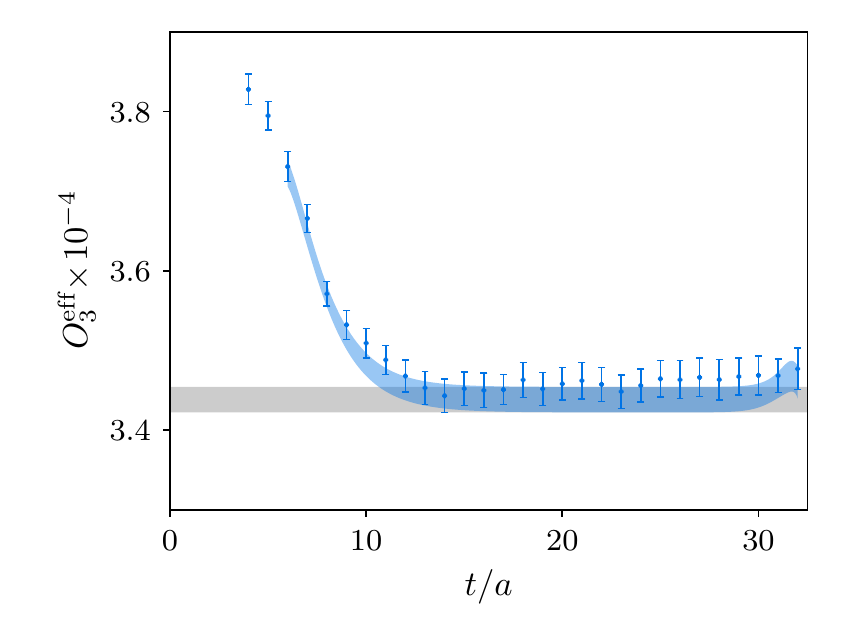}}
    \subfloat{\includegraphics{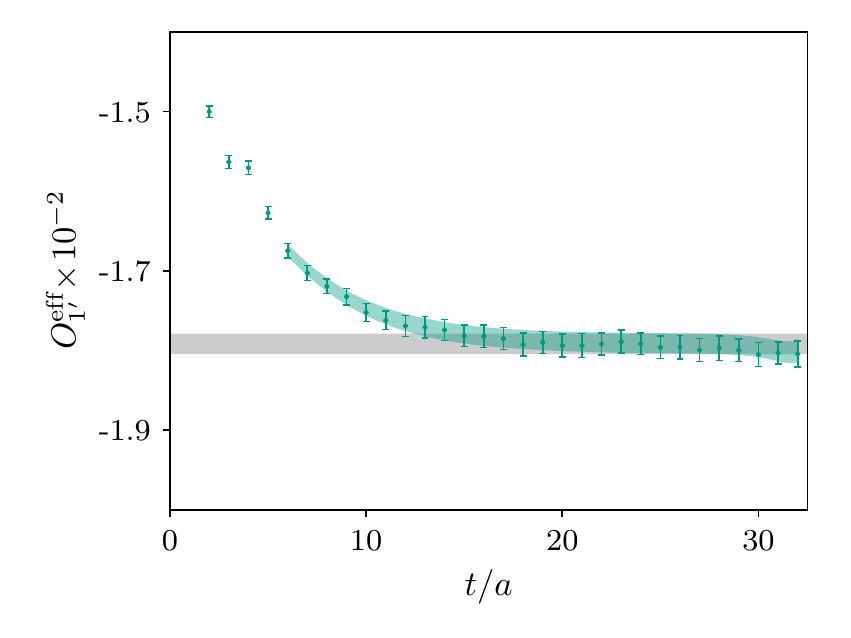}} \\
    \subfloat{\includegraphics{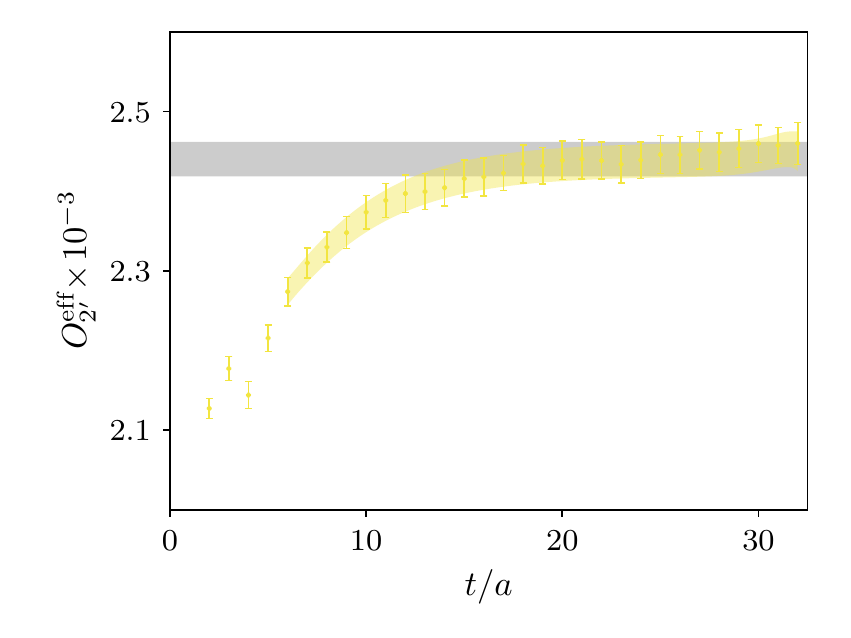}}
    \caption{Effective matrix elements, Eq.~\eqref{eq:eff_mat_elem}, for the operators $\mathcal O_k(\bm p = \bm 0)$ on the 24I, $a m_\ell = 0.01$ ensemble. The constant grey band denotes the fit results for each bare, dimensionless matrix element $a^4 \langle\pi^+ | \mathcal O_k(\bm p = \bm 0) | \pi^-\rangle$, and the colored data points and colored band denote the effective matrix element data and extrapolation band, respectively. The fit procedure is detailed in Section~\ref{sec:bare_mat_elems}.}
    \label{fig:24Iml0p01_mes}
\end{figure*}

\begin{figure*}[!htp]
    \centering
    \subfloat{\includegraphics{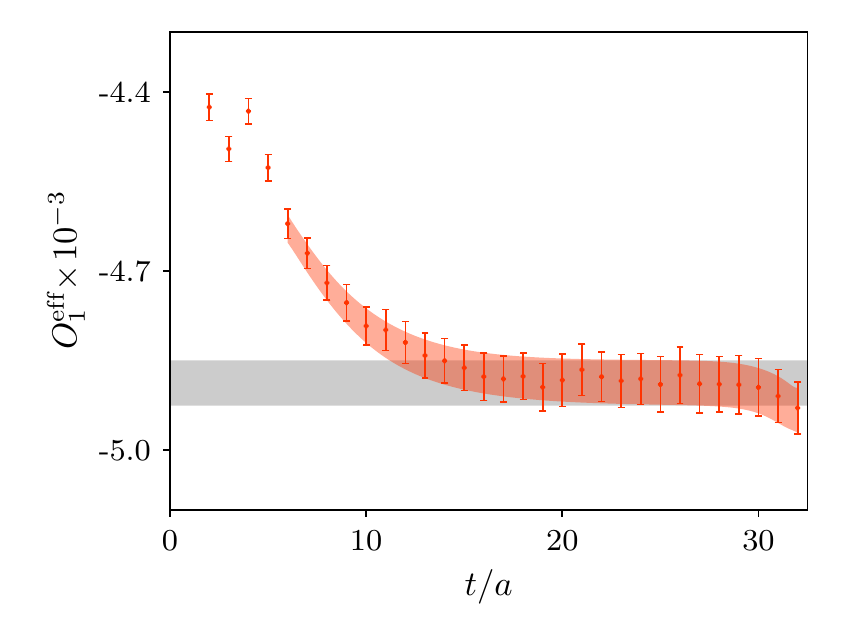}}
    \subfloat{\includegraphics{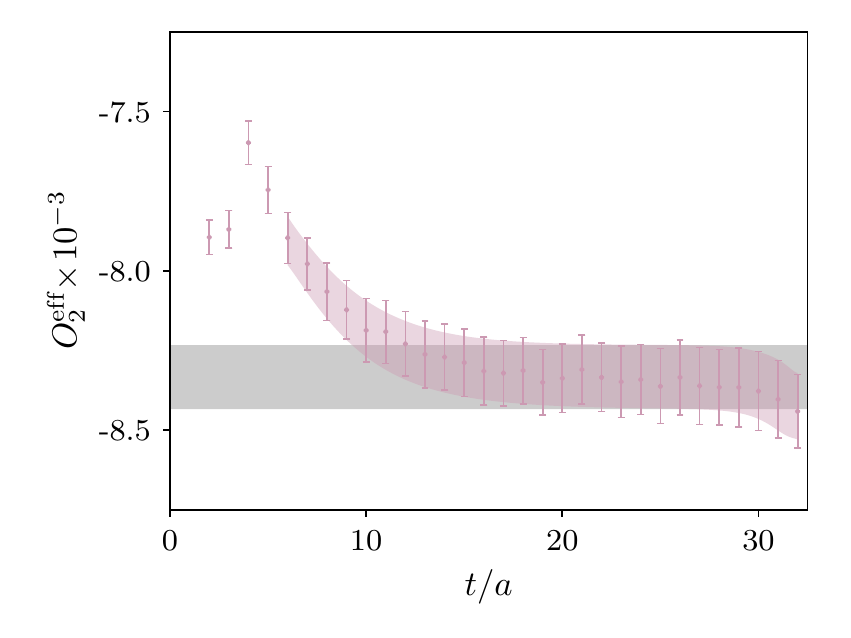}} \\
    \subfloat{\includegraphics{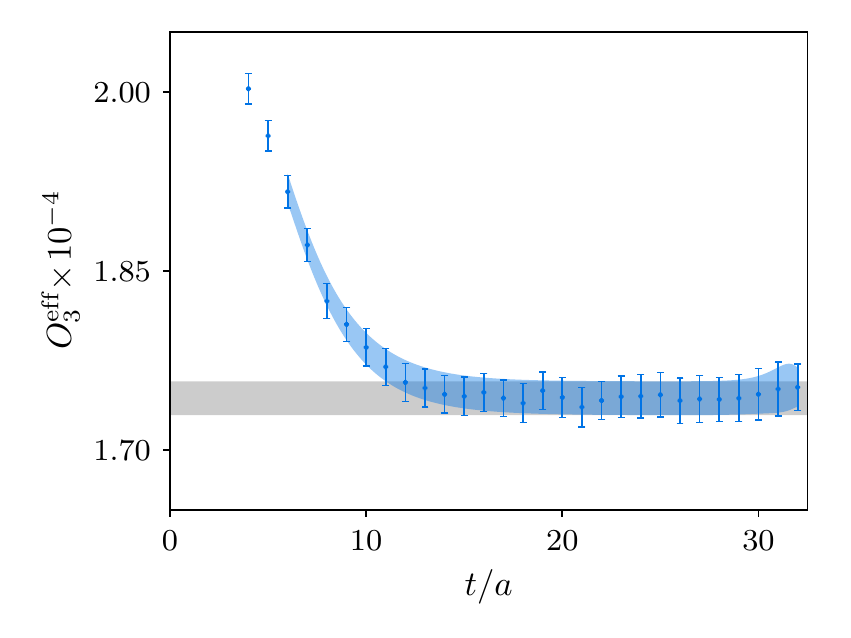}}
    \subfloat{\includegraphics{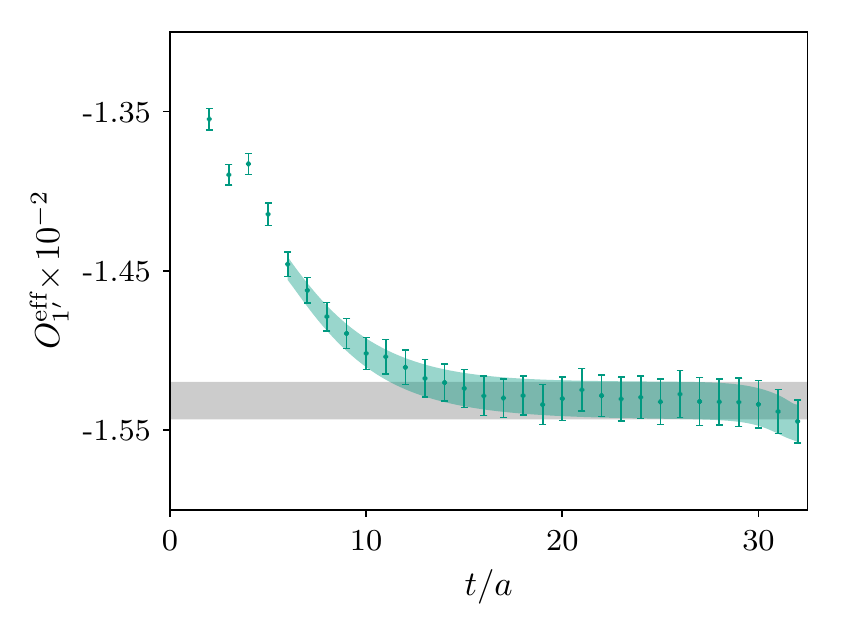}} \\
    \subfloat{\includegraphics{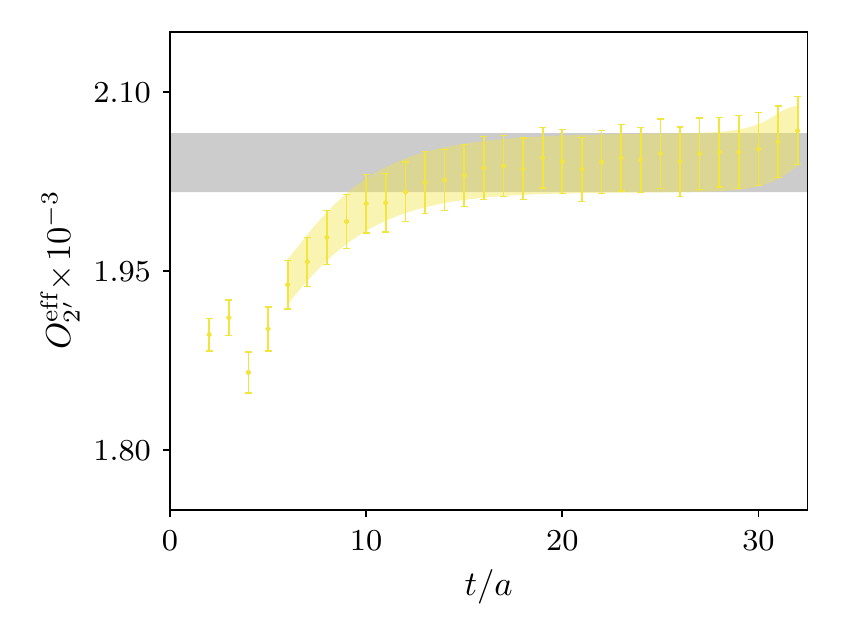}}
    \caption{As in Fig.~(\ref{fig:24Iml0p01_mes}), for the 24I, $a m_\ell = 0.005$ ensemble.}
    \label{fig:24Iml0p005_mes}
\end{figure*}

\begin{figure*}[!htp]
    \centering
    \subfloat{\includegraphics{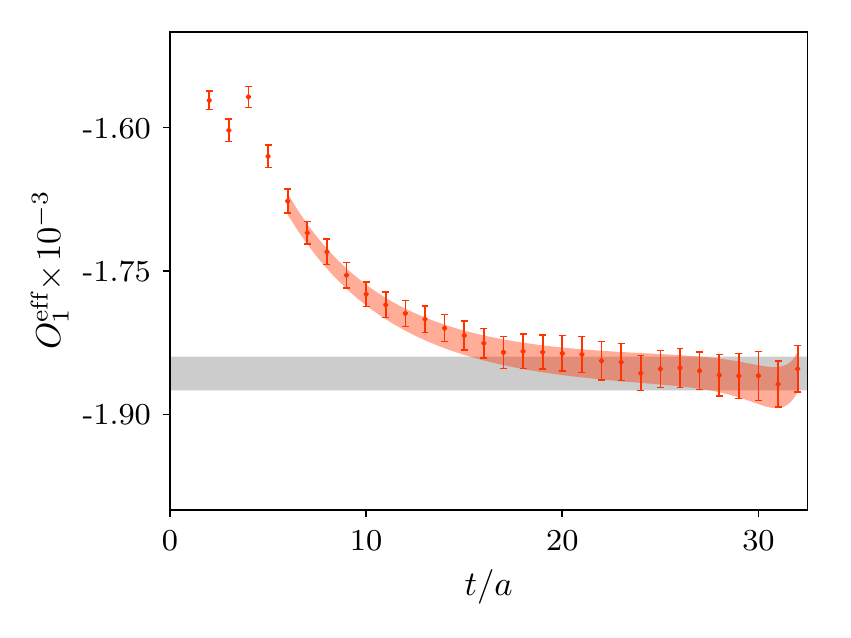}}
    \subfloat{\includegraphics{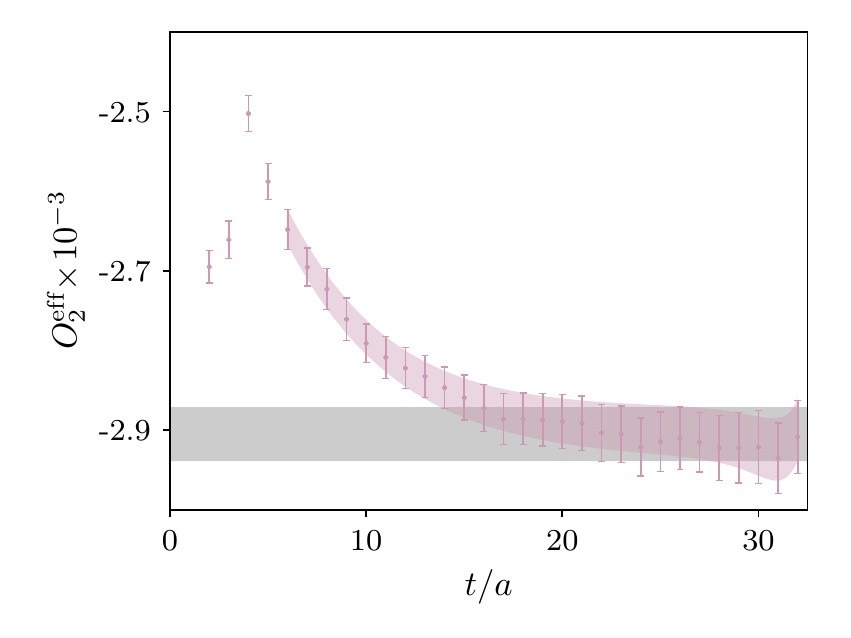}} \\
    \subfloat{\includegraphics{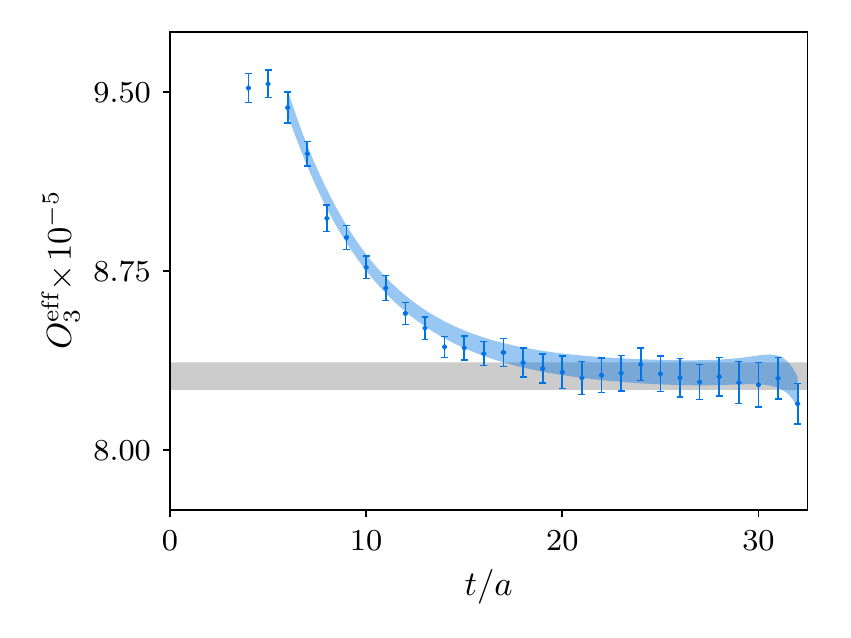}}
    \subfloat{\includegraphics{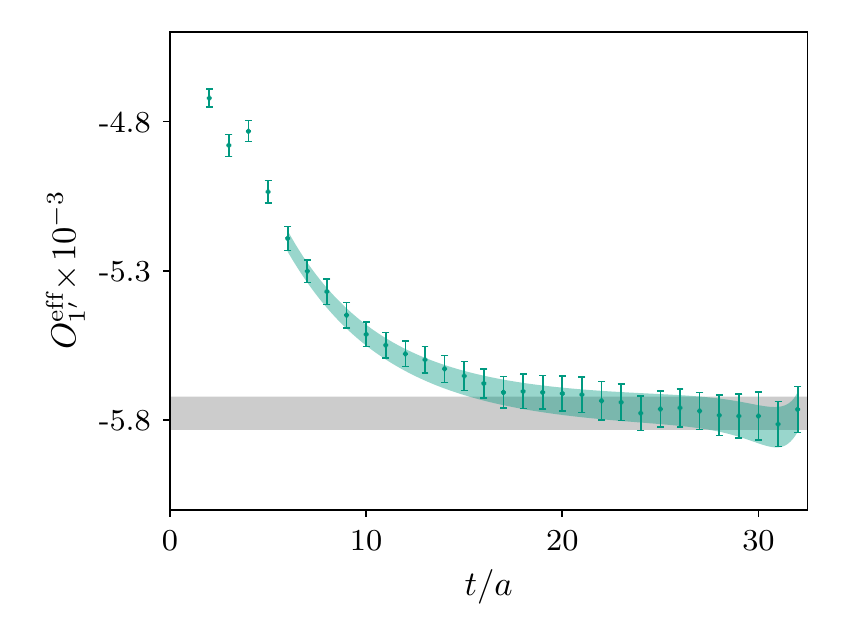}} \\
    \subfloat{\includegraphics{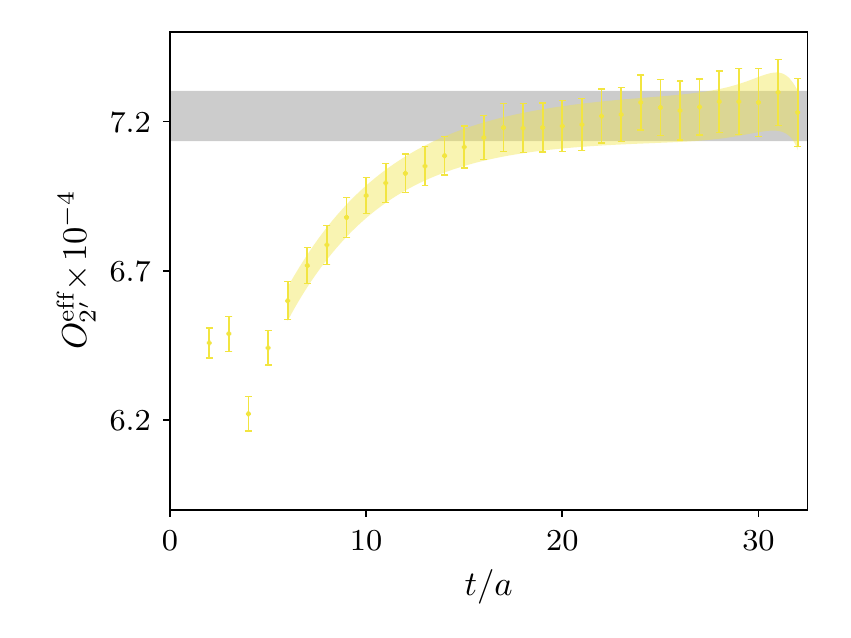}}
    \caption{As in Fig.~(\ref{fig:24Iml0p01_mes}), for the 32I, $a m_\ell = 0.008$ ensemble.}
    \label{fig:32Iml0p008_mes}
\end{figure*}

\begin{figure*}[!htp]
    \centering
    \subfloat{\includegraphics{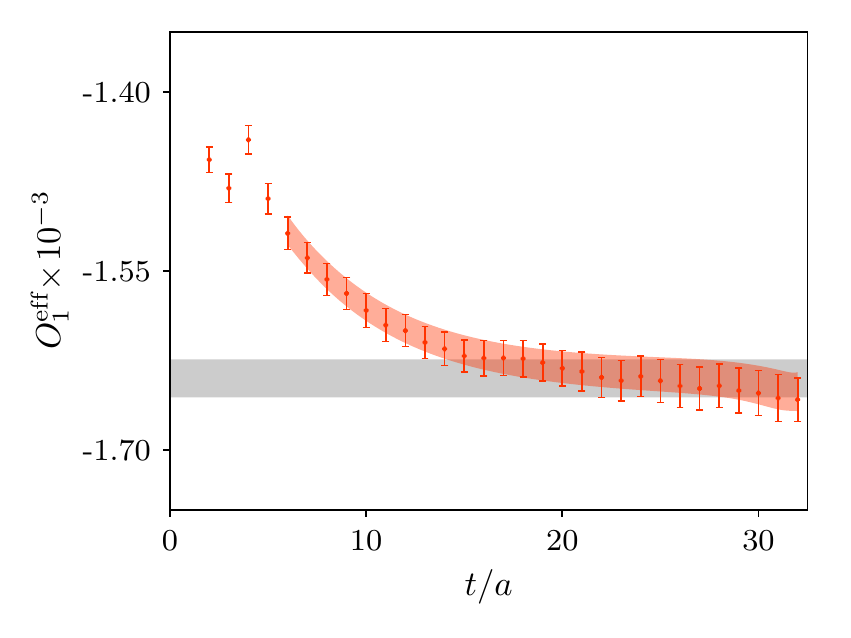}}
    \subfloat{\includegraphics{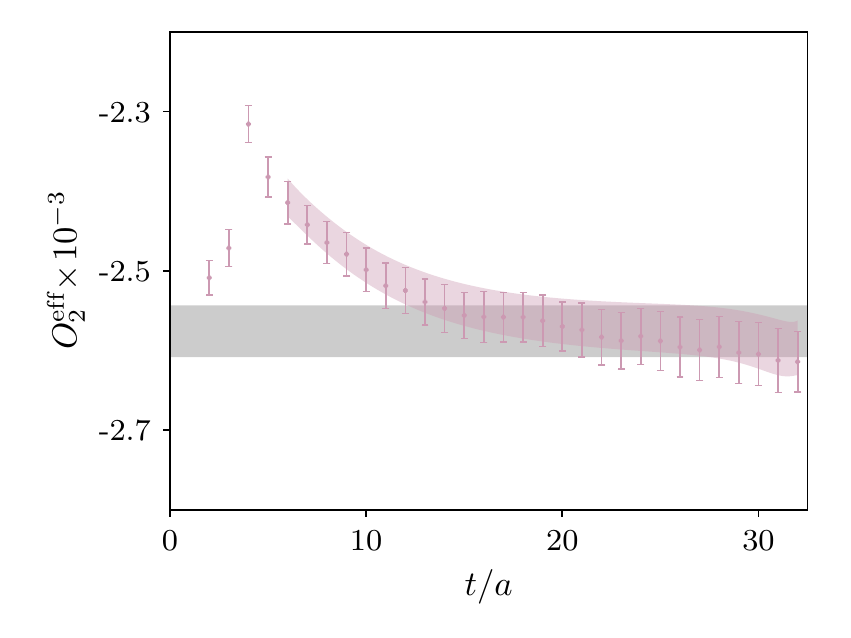}} \\
    \subfloat{\includegraphics{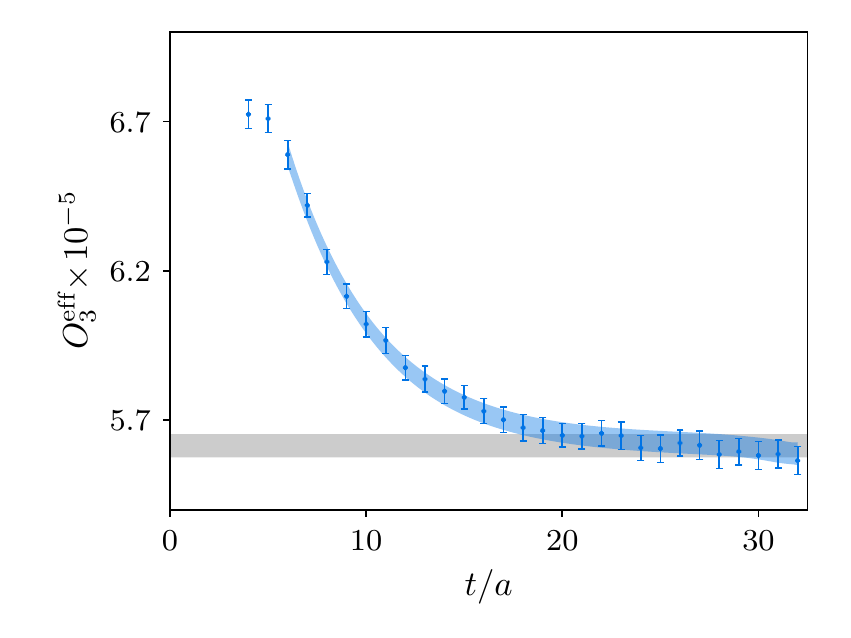}}
    \subfloat{\includegraphics{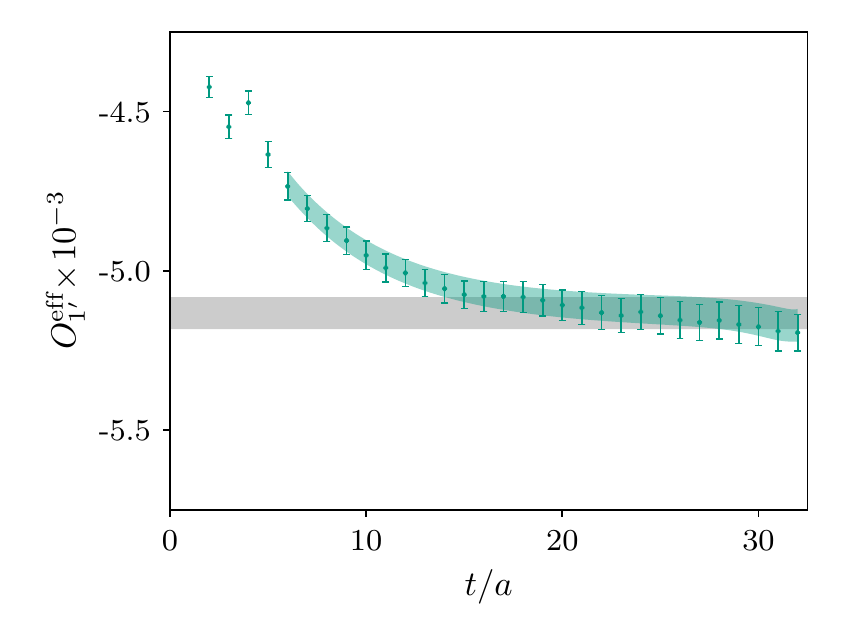}} \\
    \subfloat{\includegraphics{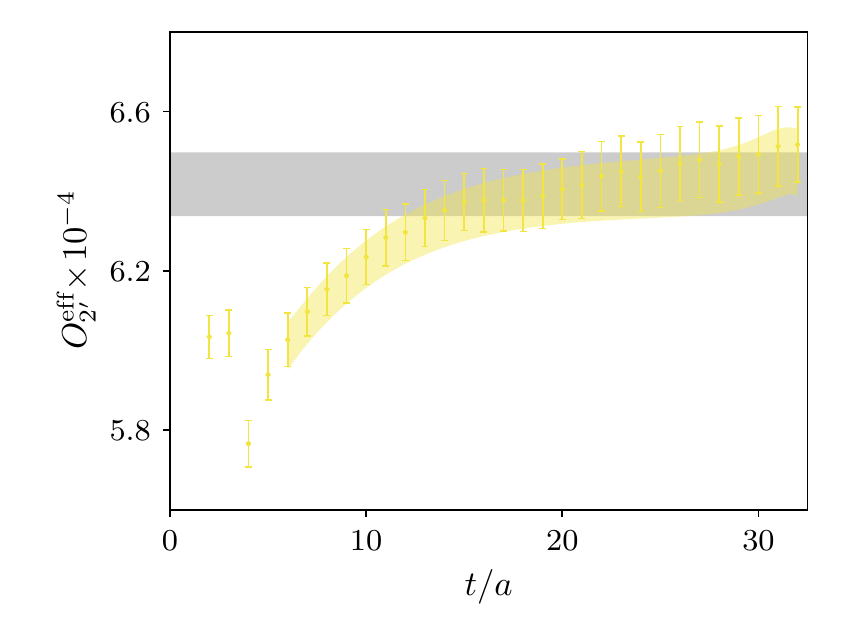}}
    \caption{As in Fig.~(\ref{fig:24Iml0p01_mes}), for the 32I, $a m_\ell = 0.006$ ensemble.}
    \label{fig:32Iml0p006_mes}
\end{figure*}

Figs.~(\ref{fig:24Iml0p01_mes})-(\ref{fig:32Iml0p006_mes}) display the remaining fits to the effective matrix elements (Eq.~\eqref{eq:eff_mat_elem}) that were not depicted in Fig.~\eqref{fig:ex_r_ratio_fits}. The fit procedure is described in Section~\ref{sec:bare_mat_elems} of the main text. The number of gauge field configurations per ensemble used in each matrix element extraction, $n_\mathrm{cfgs}$, and the corresponding bare matrix elements in lattice units, Eq.~\eqref{eq:mat_elems_latt_units}, are shown in Table~\ref{table:ratio_fits}.

\begin{table*}[!htp]
    \setlength{\tabcolsep}{5pt}
    \centering
    \begin{tabular}{ cc | c | ccccc } \hline \hline \rule{0cm}{0.4cm}Ensemble & $a m_\ell$ & $n_\mathrm{cfgs}$ & $a^4 \langle \pi^+ | \mathcal{O}_1 | \pi^- \rangle$ & $a^4 \langle \pi^+ | \mathcal{O}_2 | \pi^- \rangle$ & $a^4 \langle \pi^+ | \mathcal{O}_3 | \pi^- \rangle$ & $a^4 \langle \pi^+ | \mathcal{O}_{1'} | \pi^- \rangle$ & $a^4 \langle \pi^+ | \mathcal{O}_{2'} | \pi^- \rangle$ \\  \hline \rule{0cm}{0.4cm}\multirow{2}{*}{24I} & 0.01 & 52 & -0.005804(41) & -0.010023(91) & 0.0003442(16) & -0.01794(13) & 0.002445(22) \\ 
    & 0.005 & 53 & -0.004891(38) & -0.00834(11) & 0.0001742(14) & -0.01533(12) & 0.002043(26) \\ 
    \hline \rule{0cm}{0.4cm}\multirow{3}{*}{32I} & 0.008 & 33 &  -0.001862(17) & -0.002917(34) & 0.00008286(58) & -0.005791(53) & 0.0007248(86) \\ 
    & 0.006 & 42 & -0.001644(16) & -0.002587(36) & 0.00005600(40) & -0.005145(50) & 0.0006445(87) \\
    & 0.004 & 47 & -0.001482(15) & -0.002331(31) & 0.00003391(40) & -0.004669(47) & 0.0005822(78) \\
    \hline \hline \end{tabular}
    \caption{Determination of bare matrix elements $a^4\langle\pi^+ | \mathcal O_k(\bm p = \bm 0)|\pi^-\rangle$ on each ensemble for each operator $\mathcal O_k(x)$ in the BSM basis, Eq.~\eqref{eq:short_distance_operators}, extracted from fits to the effective matrix elements (Eq.~\eqref{eq:eff_mat_elem}) as described in the text. The effective matrix elements are computed on $n_\mathrm{cfgs}$ configurations on the respective ensemble.}
    \label{table:ratio_fits}
\end{table*}

\section{Vector and axial-vector renormalization coefficients}
\label{appendix:vector_axial_rcs}

Calculation of the scale and scheme-independent vector and axial-vector-current renormalization coefficients $\mathcal{Z}_j(a)$, with $j\in \{V, A\}$, proceeds through the vector (Eq.~\eqref{eq:vector_correlator}) and axial-vector three-point functions,
\begin{equation}
    G_A^\mu(q; a, m_\ell) = \frac{1}{V}\sum_{x, x_1, x_2} e^{i (p_1\cdot x_1 - p_2\cdot x_2  + q\cdot x)} \langle 0 | u(x_1) A^\mu(x) \overline d(x_2) |0\rangle,
    \label{eq:vector_axial_correlator}
\end{equation}
where $A^\mu(x) = \overline{u}(x)\gamma^\mu\gamma_5 d(x)$. The momenta $p_1, p_2,$ and $q$ are subject to the symmetric constraint, Eq.~\eqref{eq:sym_constraint}, and parameterized identically to the modes used in the calculation of the four-quark operator renormalizations (Eq.~\eqref{eq:mom_modes}) with $k\in \{2, 3, 4, 5\}$. The lattice spacing dependence is made explicit in this section. The amputated three-point functions
\begin{equation}
    \Lambda_{j}^\mu(q; a, m_\ell) = S^{-1}(p_1; a, m_\ell) G_{j}^\mu (q; a, m_\ell) S^{-1}(p_2; a, m_\ell),
\end{equation}
with $j\in \{V, A\}$, are used to compute the renormalization coefficients,
\begin{align}\begin{split}
    \frac{1}{12\tilde q^2} \frac{\mathcal Z_V(\mu^2; a, m_\ell)}{\mathcal Z_q^{\mathrm{RI/sMOM}}(\mu^2; a, m_\ell)} \mathrm{Tr}\left[ \tilde q_\mu \Lambda_V^\mu(q; a, m_\ell) \slashed{\tilde q} \right] \bigg|_{\mathrm{sym}} &= 1, \\
    \frac{1}{12\tilde q^2} \frac{\mathcal Z_A(\mu^2; a, m_\ell)}{\mathcal Z_q^{\mathrm{RI/sMOM}}(\mu^2; a, m_\ell)} \mathrm{Tr}\left[ \tilde q_\mu \Lambda_A^\mu(q; a, m_\ell) \gamma_5 \slashed{\tilde q} \right] \bigg|_{\mathrm{sym}} &= 1,
    \label{eq:vector_axial_renorm}
\end{split}\end{align}
where $\tilde p_\mu = \frac{2}{a} \sin (\frac{a}{2} p_\mu)$ is the lattice momentum. Note that the quark-field renormalization in Eq.~\eqref{eq:vector_axial_renorm} is defined in the RI/sMOM scheme~\cite{Sturm:2009kb},
\begin{equation}
    \mathcal Z_q^{\mathrm{RI/sMOM}}(\mu^2; a, m_\ell)\bigg|_{p^2 = \mu^2} = \frac{i}{12 \tilde p^2} \mathrm{Tr}[S^{-1}(p; a, m_\ell)\slashed{\tilde p}]\bigg|_{p^2 = \mu^2}, \label{eq:quark_renorm_RIsMOM}
\end{equation}
which differs from the RI$\gamma$ scheme~\cite{Boyle:2017skn} of Eq.~\eqref{eq:quark_renorm}; $\mathcal Z_V$ and $\mathcal Z_A$ are scheme-independent, hence may be computed in any scheme. The chiral limits $\mathcal Z_V(\mu^2; a)$ and $\mathcal Z_A(\mu^2; a)$ of $\mathcal Z_V(\mu^2; a, m_\ell)$ and $\mathcal Z_A(\mu^2; a, m_\ell)$ are evaluated by a joint, correlated linear extrapolation of $\{\mathcal Z_q^{\mathrm{RI/sMOM}}, \mathcal Z_V, \mathcal Z_A\}$ in $m_\ell$, identical to the procedure used in the $am_\ell\rightarrow 0$ extrapolation of $\{\mathcal Z_q^{\mathrm{RI}\gamma} / \mathcal Z_V, F_{nm}\}$, as described in Section~\ref{sec:renorm} of the text (Eqs.~\eqref{eq:Fmn_dfn}-\eqref{eq:amell_extrap_model}). 

\begin{figure*}[!htp]
    \centering
    \subfloat{\includegraphics{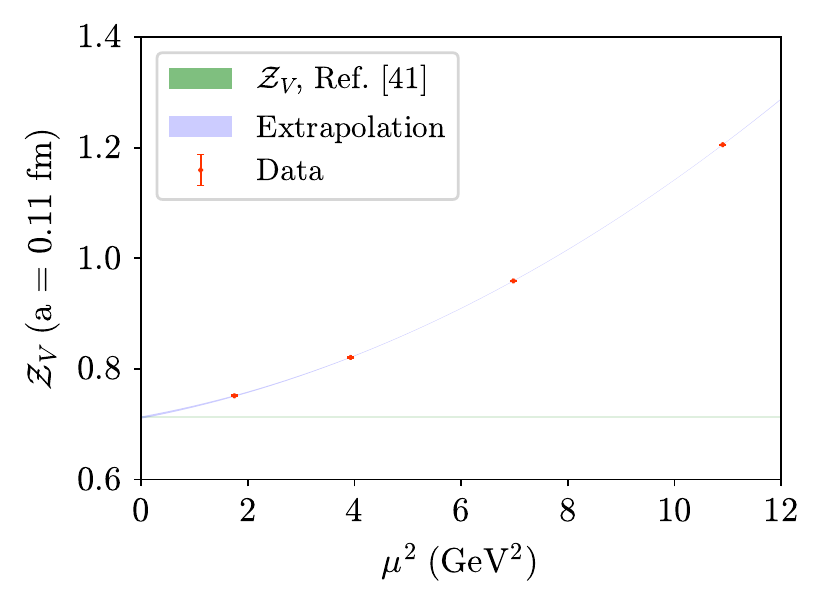}}
    \subfloat{\includegraphics{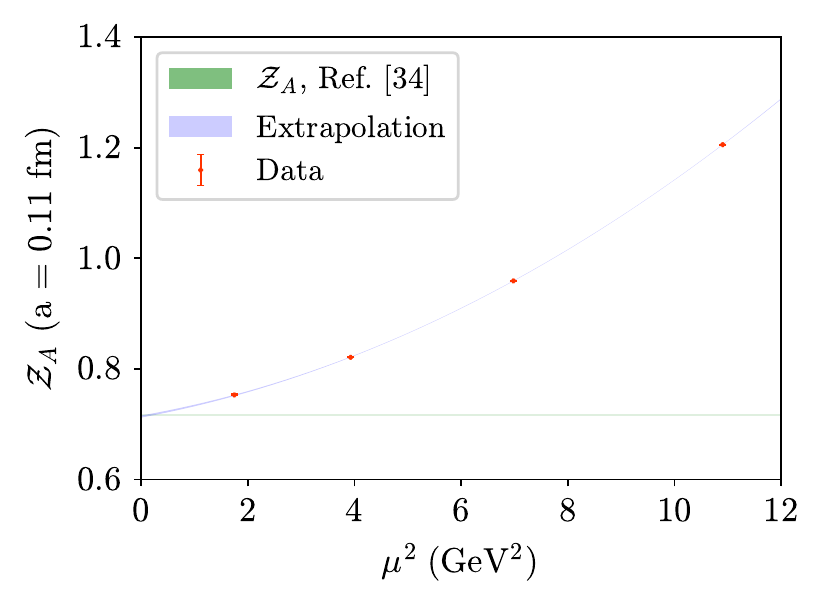}} \\
    \subfloat{\includegraphics{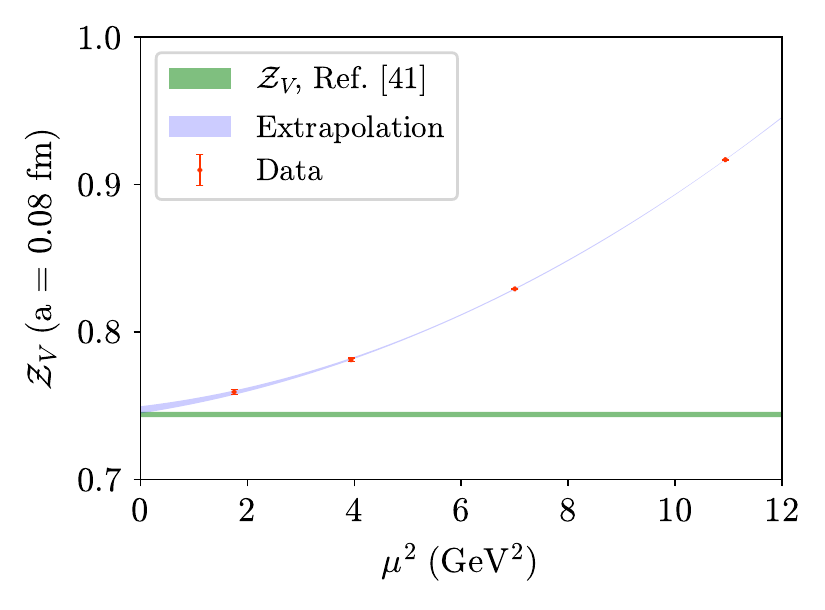}}
    \subfloat{\includegraphics{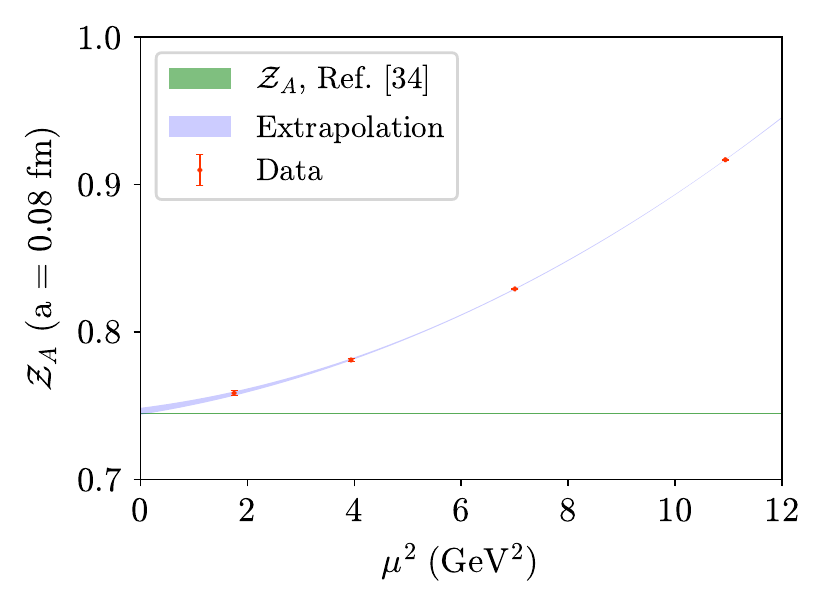}}
    \caption{Vector and axial-vector renormalization coefficients computed by the procedure described in the text, and extrapolated to $\mu^2 = 0$ with the model given in Eq.~\eqref{eq:vector_axial_model}. The red data points are the computed data, Eq.~\eqref{eq:vector_axial_renorm}, the blue bands show the extrapolation to $\mu^2\rightarrow 0$, and the green bands denote the chiral limit value of $\mathcal Z_A$ and $\mathcal Z_V$ computed in Refs.~\cite{Detmold:2020jqv,PhysRevD.93.074505}.}
    \label{fig:axial_vector_renorm}
\end{figure*}

Although the renormalization coefficients $\mathcal Z_V$, $\mathcal Z_A$ are scale-independent, the RI procedure introduces scale-dependence from the kinematic setup (Eq.~\eqref{eq:sym_constraint}). This scale-dependence is removed by fitting $\mathcal Z_j(\mu^2; a)$ to a power series in $\mu^2$ and taking the $\mu^2\rightarrow 0$ limit as described in Ref.~\cite{PhysRevLett.126.202001}, with fit model:
\begin{equation}
    \mathcal Z_j(\mu^2; a) = \mathcal Z_j(a) + c_j^{(1)}(a) \mu^2 + c_j^{(2)}(a) \mu^4.
    \label{eq:vector_axial_model}
\end{equation}
Here $\mathcal Z_j(a), c_j^{(1)}(a)$, and $c_j^{(2)}(a)$ are coefficients which are determined by correlated $\chi^2$ minimization. The fits are shown in Fig.~\eqref{fig:axial_vector_renorm}. The fits have $\chi^2/\mathrm{dof}$ ranging between 0.15 and 0.71. The best-fit value of $\mathcal Z_j(a)$ is the value that is taken for the renormalization factor, and it is determined that
\begin{align} \begin{split}
    \mathcal Z_V(0.11\;\mathrm{fm}) = 0.7119(20) \hspace{4.0cm} & \mathcal Z_V(0.08\;\mathrm{fm}) = 0.7472(24) \\
    \mathcal Z_A(0.11\;\mathrm{fm}) = 0.7137(19) \hspace{4.0cm} &\mathcal Z_A(0.08\;\mathrm{fm}) = 0.7462(23). \label{eq:axial_vector_rcs}
\end{split} \end{align}
The results show that $\mathcal Z_V = \mathcal Z_A$ within statistical precision as expected. The determination presented in this work is consistent with the determination of $\mathcal Z_V$ in Ref.~\cite{PhysRevD.93.074505} for the $a = 0.08\;\mathrm{fm}$ and $a = 0.11\;\mathrm{fm}$ ensembles, and with $\mathcal Z_A$ in Ref.~\cite{Detmold:2020jqv} for the $a = 0.08\;\mathrm{fm}$ ensembles, although $\mathcal Z_A$ differs from the $a = 0.11\;\mathrm{fm}$ value in that work by about one standard deviation. This deviation may be due to discrepancies in the procedure used to extract $\mathcal Z_A$, as the fit model (Eq.~\eqref{eq:vector_axial_model}) does not capture all the discretization artifacts present in the data.

\clearpage
\section{Renormalization coefficient $am_\ell\rightarrow 0$ extrapolation}
\label{app:amell_extrapolation}

\begin{figure*}[!htp]
    \centering
    \subfloat{\includegraphics{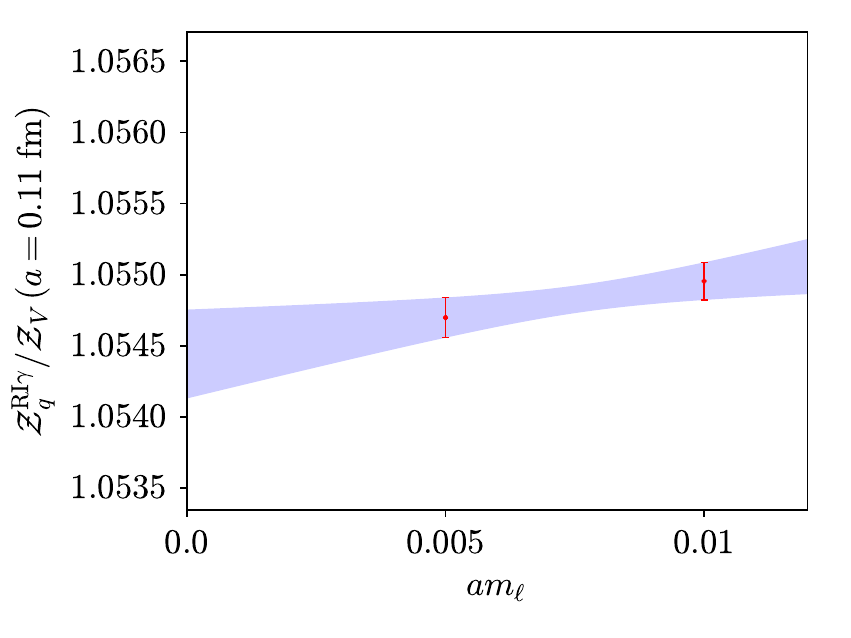}} 
    \caption{The $am_\ell\rightarrow 0$ extrapolation for the RI quark-field renormalization $\mathcal Z_q^{\mathrm{RI}\gamma} / \mathcal{Z}_V$, Eq.~\eqref{eq:quark_renorm}, computed on the $a = 0.11\;\mathrm{fm}$ ensembles at $q = \frac{2\pi}{aL}(4, 4, 0, 0)$ and extrapolated to the chiral limit via a joint correlated linear extrapolation in $am_\ell$ (Eq.~\eqref{eq:amell_extrap_model}). The data is depicted in red, and the shaded band denotes the extrapolation.}
    \label{fig:24I_amell_ZqVA}
\end{figure*}

\begin{figure*}[!htp]
    \centering
    \subfloat{\includegraphics{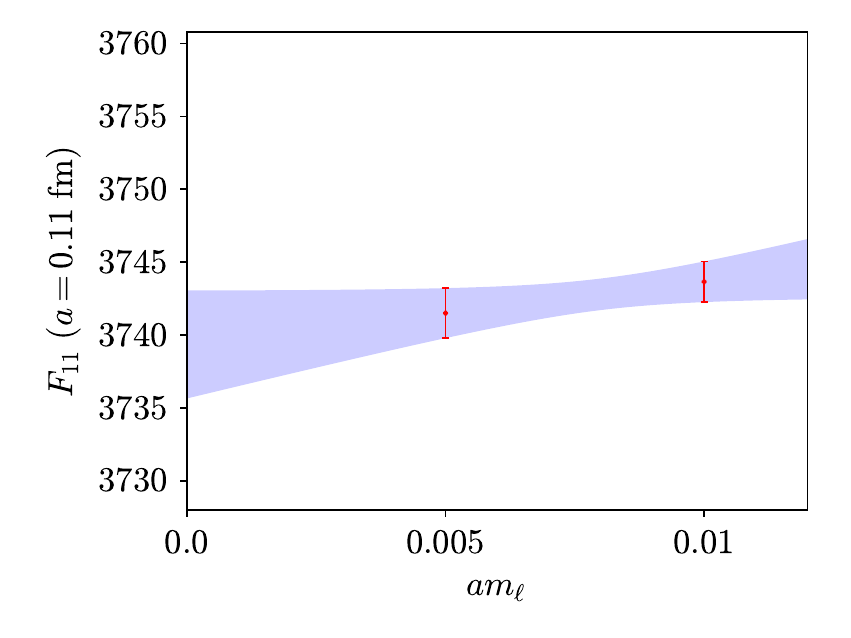}}
    \caption{As in Fig.~(\ref{fig:24I_amell_ZqVA}), $am_\ell\rightarrow 0$ extrapolation for $F_{nm}$ on the first irreducible chiral subspace $\{F_{11}\}$, for the $a = 0.11\;\mathrm{fm}$ ensembles.}
    \label{fig:24I_amell_F11}
\end{figure*}

\begin{figure*}[!htp]
    \centering
    \subfloat{\includegraphics{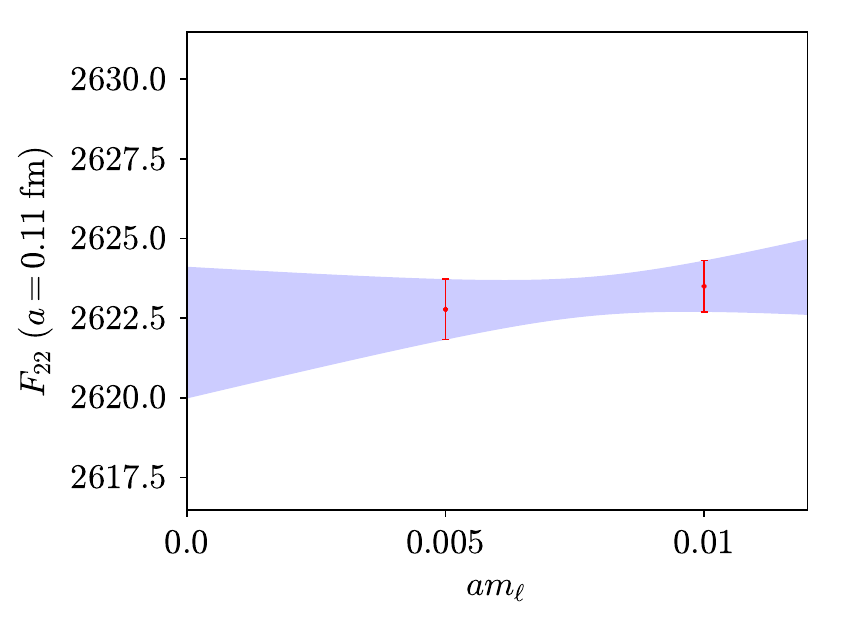}}
    \subfloat{\includegraphics{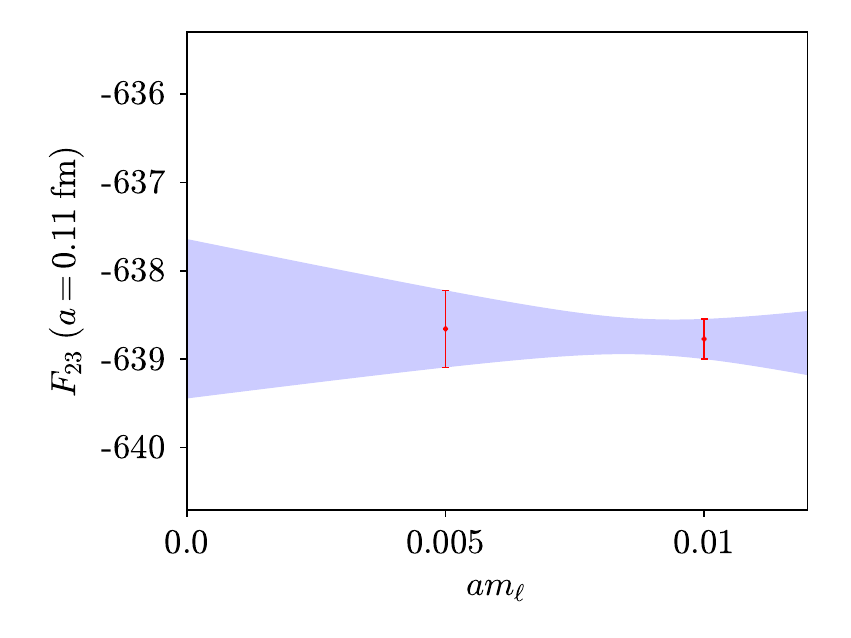}} \\
    \subfloat{\includegraphics{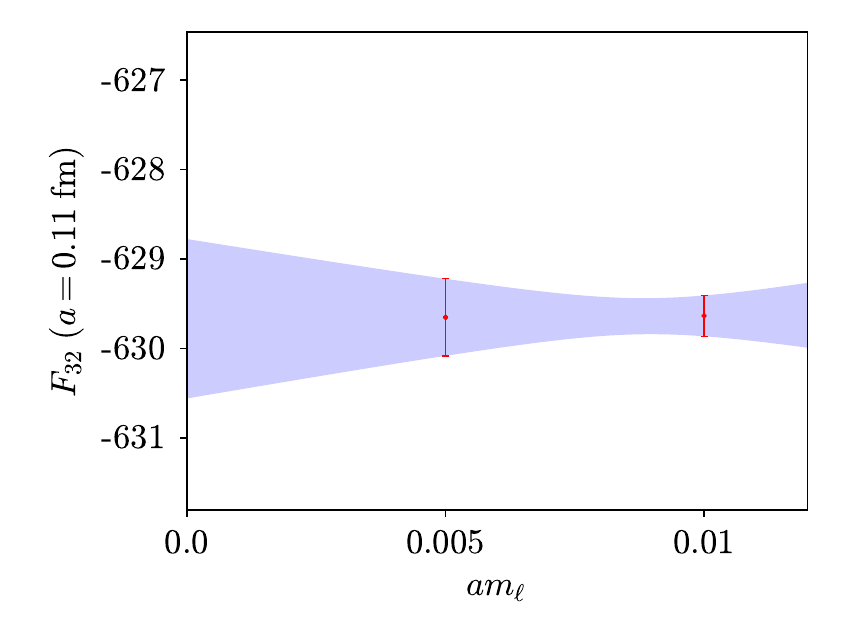}}
    \subfloat{\includegraphics{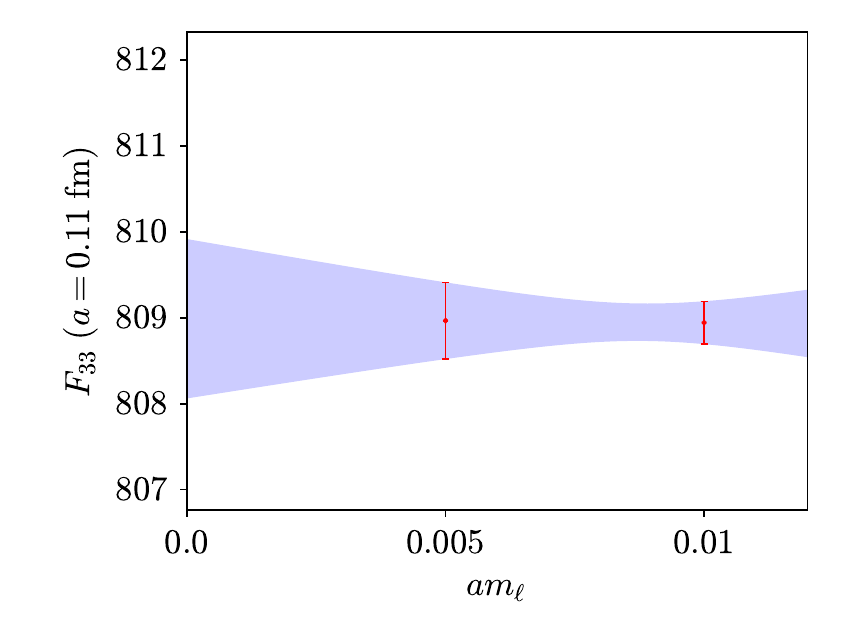}}
    \caption{As in Fig.~(\ref{fig:24I_amell_ZqVA}), $am_\ell\rightarrow 0$ extrapolation for $F_{nm}$ on the second irreducible chiral subspace $\{F_{22}, F_{23}, F_{32}, F_{33}\}$, for the $a = 0.11\;\mathrm{fm}$ ensembles.}
    \label{fig:24I_amell_F22_23_32_33}
\end{figure*}

\begin{figure*}[!htp]
    \centering
    \subfloat{\includegraphics{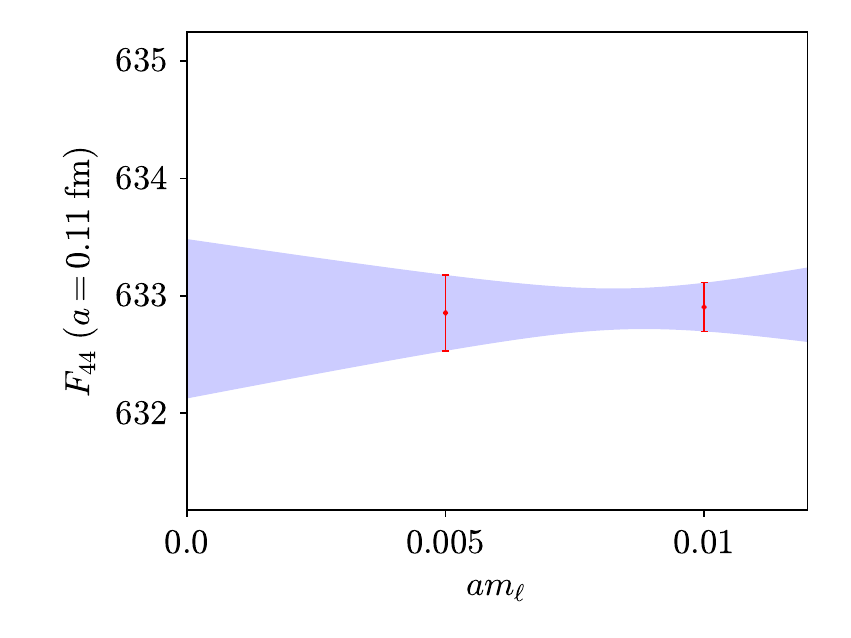}}
    \subfloat{\includegraphics{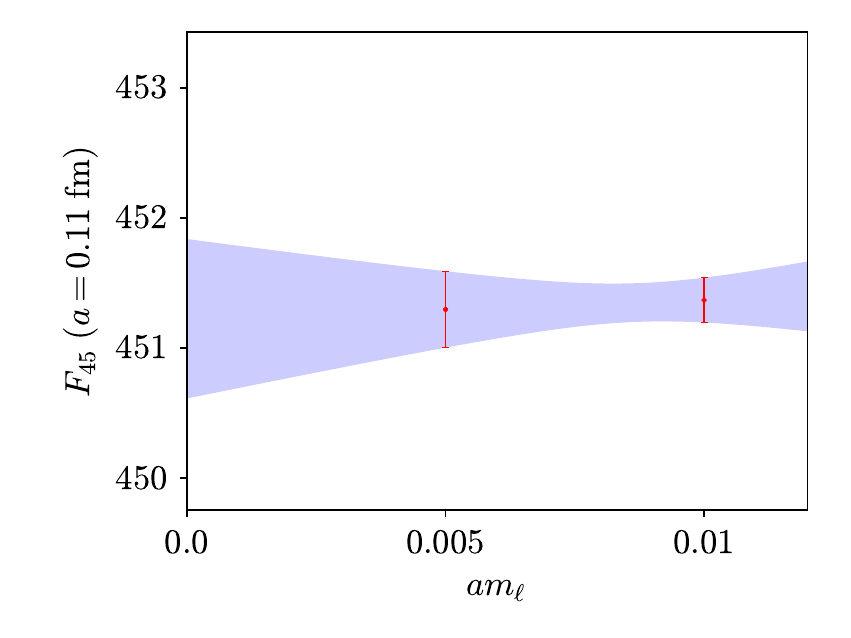}} \\
    \subfloat{\includegraphics{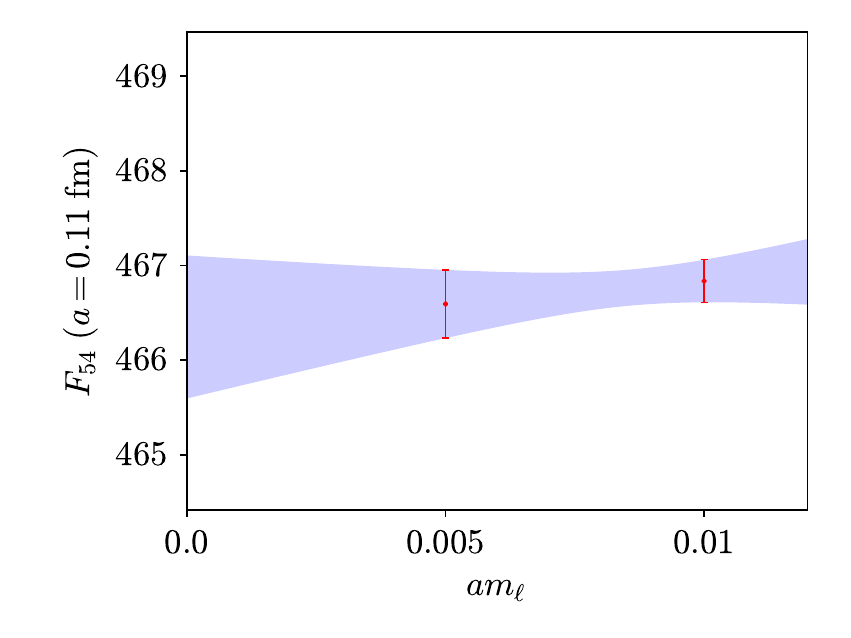}}
    \subfloat{\includegraphics{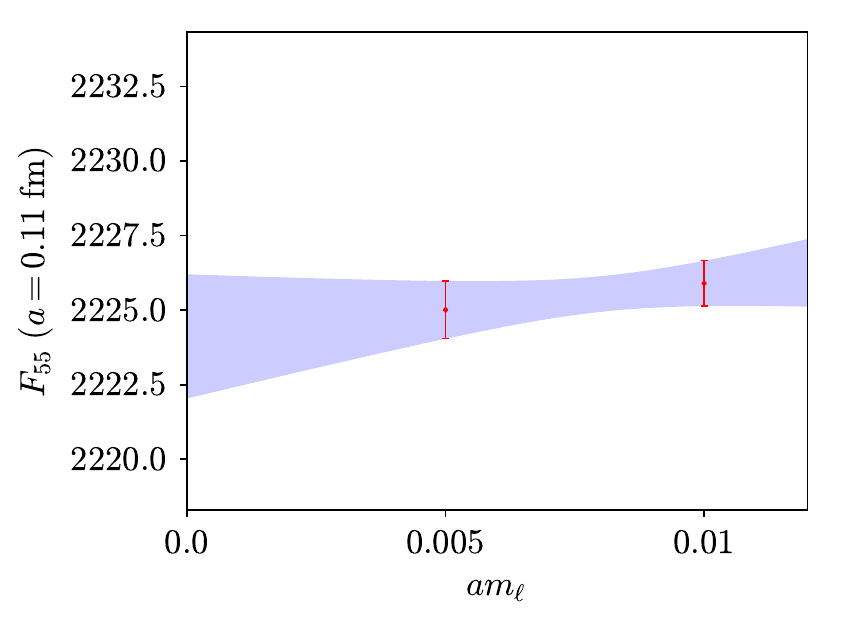}}
    \caption{As in Fig.~(\ref{fig:24I_amell_ZqVA}), $am_\ell\rightarrow 0$ extrapolation for $F_{nm}$ on the third irreducible chiral subspace $\{F_{44}, F_{45}, F_{54}, F_{55}\}$, for the $a = 0.11\;\mathrm{fm}$ ensembles.}
    \label{fig:24I_amell_F44_45_54_55}
\end{figure*}

\begin{figure*}[!htp]
    \centering
    \subfloat{\includegraphics{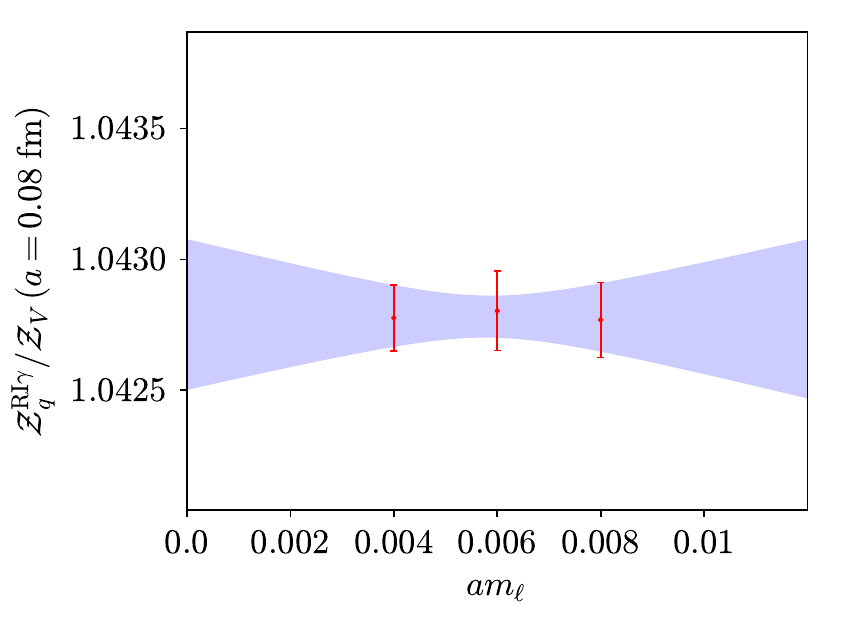}} 
    \caption{As in Fig.~(\ref{fig:24I_amell_ZqVA}), $am_\ell\rightarrow 0$ extrapolation for $\mathcal Z_q^{\mathrm{RI}\gamma} / \mathcal Z_V$, for the $a = 0.08\;\mathrm{fm}$ ensembles.}
    \label{fig:32I_amell_ZqVA}
\end{figure*}

\begin{figure*}[!htp]
    \centering
    \subfloat{\includegraphics{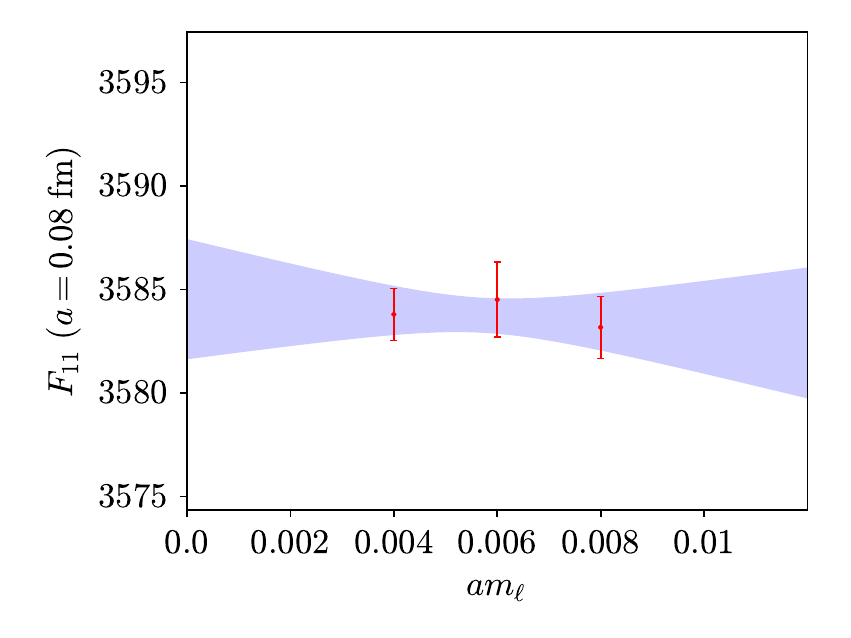}}
    \caption{As in Fig.~(\ref{fig:24I_amell_ZqVA}), $am_\ell\rightarrow 0$ extrapolation for $F_{nm}$ on the first irreducible chiral subspace $\{F_{11}\}$, for the $a = 0.08\;\mathrm{fm}$ ensembles.}
    \label{fig:32I_amell_F11}
\end{figure*}

\begin{figure*}[!htp]
    \centering
    \subfloat{\includegraphics{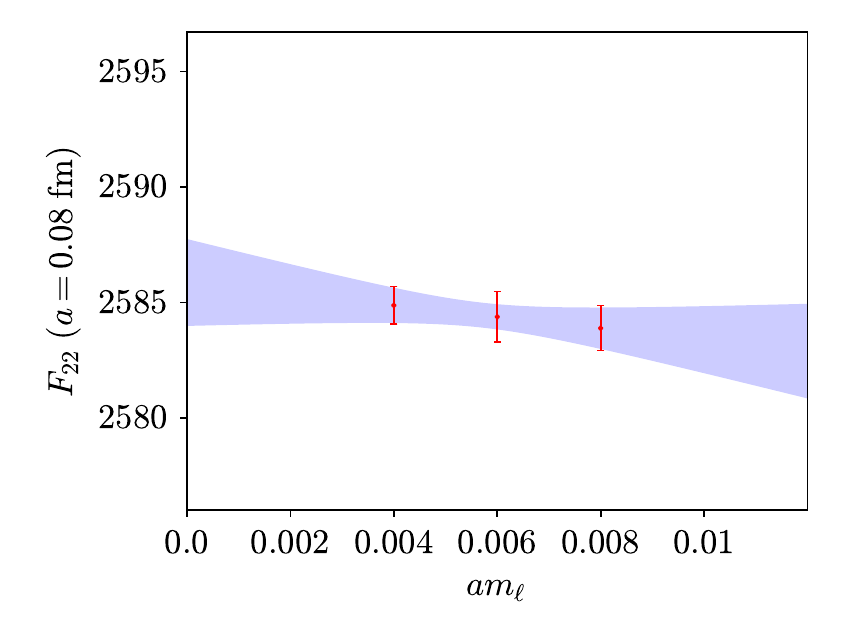}}
    \subfloat{\includegraphics{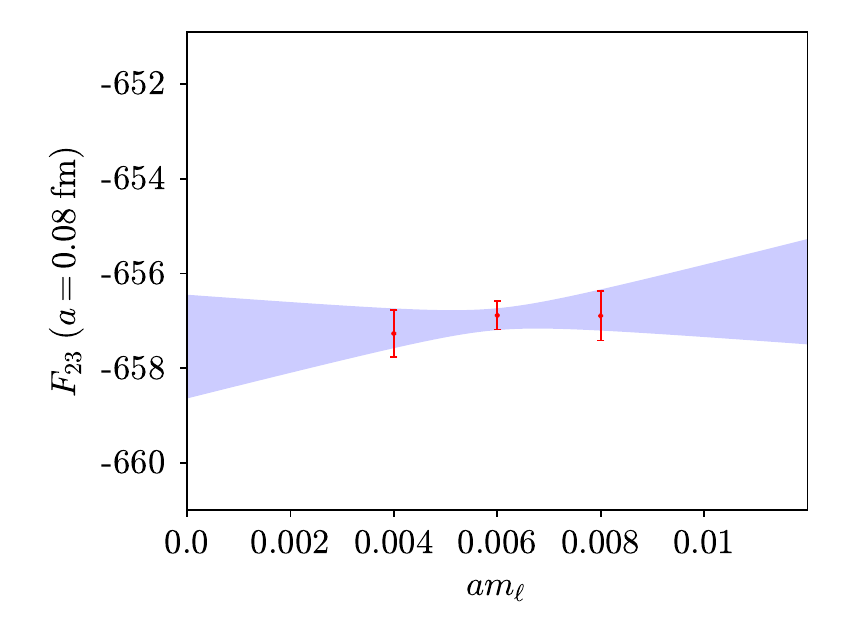}} \\
    \subfloat{\includegraphics{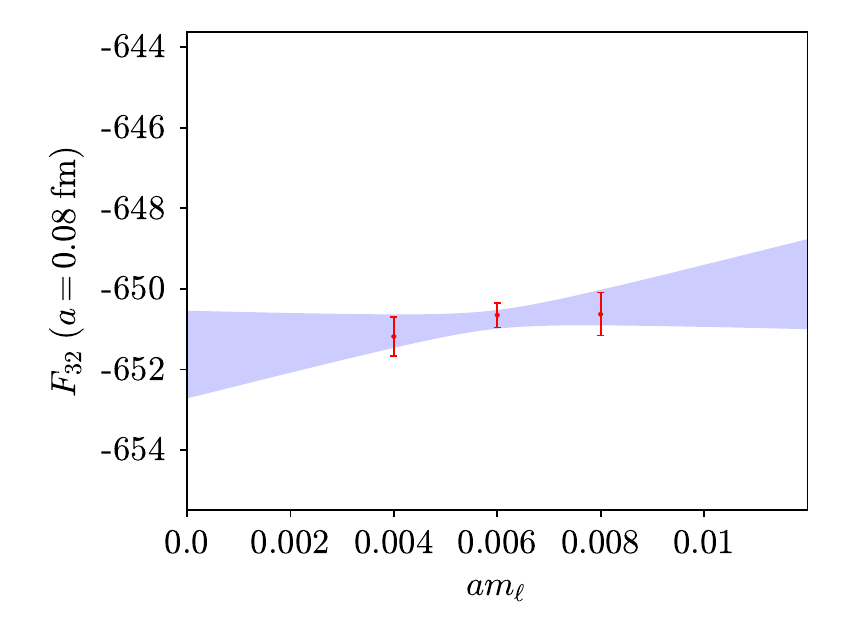}}
    \subfloat{\includegraphics{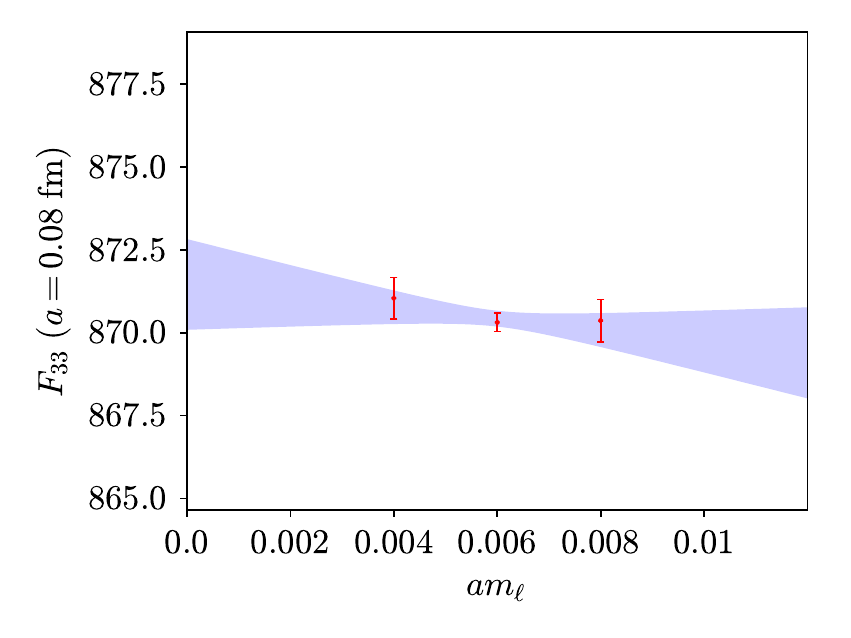}}
    \caption{As in Fig.~(\ref{fig:24I_amell_ZqVA}), $am_\ell\rightarrow 0$ extrapolation for $F_{nm}$ on the second irreducible chiral subspace $\{F_{22}, F_{23}, F_{32}, F_{33}\}$, for the $a = 0.08\;\mathrm{fm}$ ensembles.}
    \label{fig:32I_amell_F22_23_32_33}
\end{figure*}

\begin{figure*}[!htp]
    \centering
    \subfloat{\includegraphics{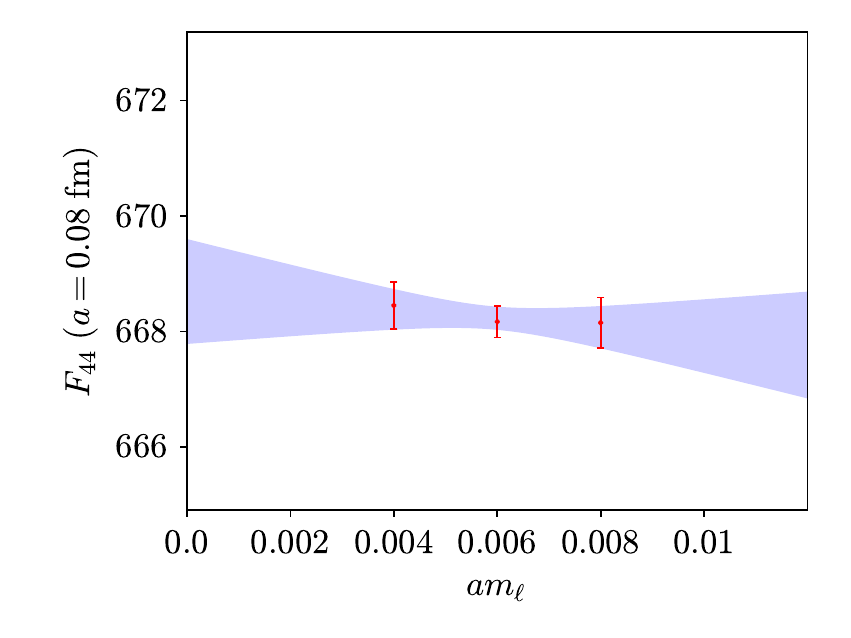}}
    \subfloat{\includegraphics{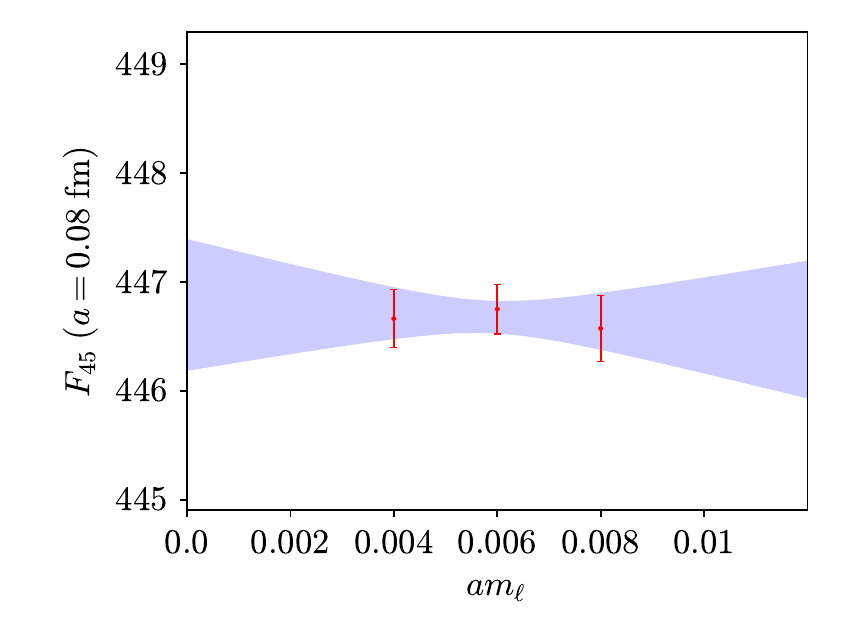}} \\
    \subfloat{\includegraphics{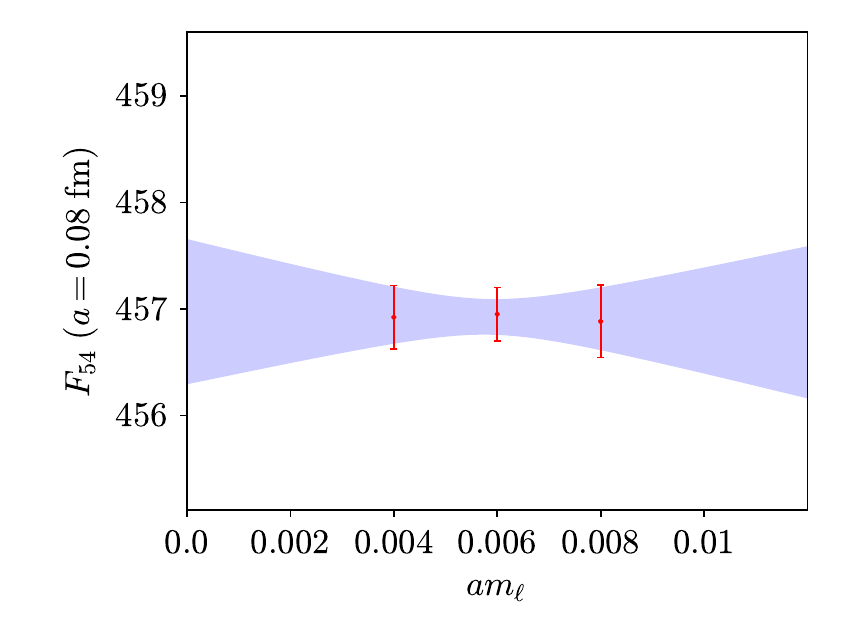}}
    \subfloat{\includegraphics{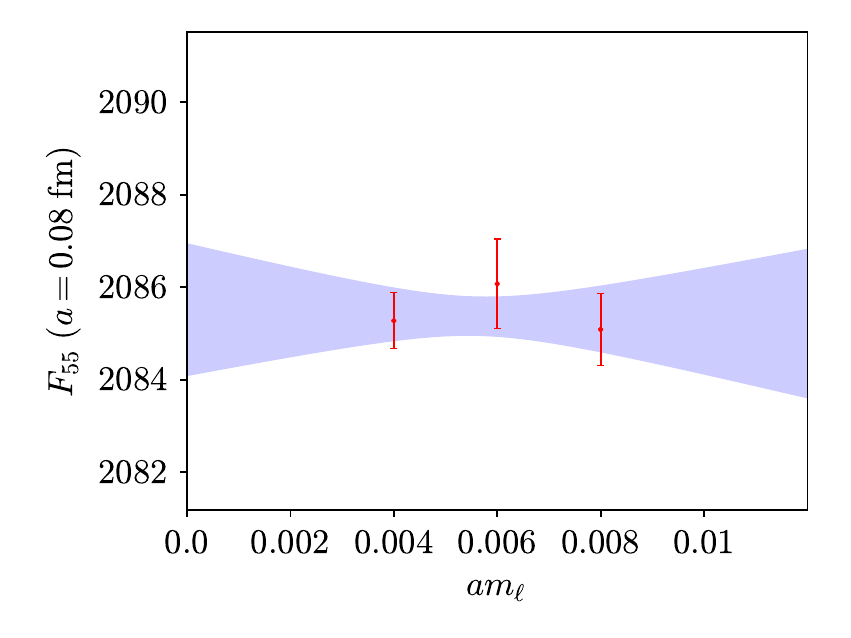}}
    \caption{As in Fig.~(\ref{fig:24I_amell_ZqVA}), $am_\ell\rightarrow 0$ extrapolation for $F_{nm}$ on the third irreducible chiral subspace $\{F_{44}, F_{45}, F_{54}, F_{55}\}$, for the $a = 0.08\;\mathrm{fm}$ ensembles.}
    \label{fig:32I_amell_F44_45_54_55}
\end{figure*}

Figs.~(\ref{fig:24I_amell_ZqVA})-(\ref{fig:32I_amell_F44_45_54_55}) display the $a m_\ell\rightarrow 0$ extrapolations of $\mathcal Z_q^{\mathrm{RI}\gamma} / \mathcal Z_V$ and $F_{nm}$, as described in Section~\ref{sec:renorm} of the text. Each renormalization coefficient is evaluated at $q = \frac{2\pi}{L}(4, 4, 0, 0)$, which is the lattice momentum corresponding to the scale $\mu = \mu_4$. In each of Figs.~(\ref{fig:24I_amell_ZqVA})-(\ref{fig:32I_amell_F44_45_54_55}), the $\mu$ dependence of $(\mathcal Z_q^{\mathrm{RI}\gamma} / \mathcal{Z}_V )(\mu^2; a)$ and the $q$ dependence of $F_{nm}(q; a)$ has been suppressed for clarity. The data is observed to have very mild dependence on $a m_\ell$. 

\end{appendices}

\end{document}